\documentclass[a4paper,reqno,twoside,11pt]{amsart}

\usepackage{amssymb}
\usepackage{amsxtra}            
\usepackage{epsfig}
\usepackage{cite}              
\usepackage{psfrag}
\usepackage{pst-tree}
\usepackage{url}

\newtheorem{result}{Theorem}    

\newtheorem{theorem}{Theorem}[section]
\newtheorem{proposition}[theorem]{Proposition}
\newtheorem{lemma}[theorem]{Lemma}
\newtheorem{corollary}[theorem]{Corollary}

\theoremstyle{definition}

\newtheorem{example}[theorem]{Example}

\theoremstyle{remark}
\newtheorem{remark}[theorem]{Remark}

\numberwithin{equation}{section}

\newcommand{\abs}[1]{\lvert#1\rvert}

\newcommand{\piste}{\,\cdot\,}

\newcommand{\average}[1]{\langle#1\rangle}

\newcommand{\eg}{\textit{e.g.}}
\newcommand{\ie}{\textit{i.e.}}

\newcommand{\Z}{\mathbb{Z}}
\newcommand{\N}{\mathbb{N}}
\newcommand{\R}{\mathbb{R}}
\newcommand{\C}{\mathbb{C}}

\newcommand{\es}{\mathbb{S}}
\newcommand{\Ham}{\mathcal{H}}

\newcommand{\bigunit}{[-1,1]}
\newcommand{\torus}{{\mathbb{T}^d}}
\newcommand{\T}{\mathcal{T}}
\newcommand{\W}{\mathcal{W}}
\newcommand{\A}{\mathcal{A}}

\newcommand{\K}{\mathcal{K}}
\newcommand{\I}{\mathcal{I}}

\newcommand{\E}{\mathcal{E}}

\newcommand{\el}{\mathcal{L}}
\newcommand{\D}{\mathcal{D}}
\newcommand{\order}[1]{\mathcal{O}(#1)}

\newcommand{\inv}{^{-1}}

\DeclareMathOperator{\sgn}{sgn}
\DeclareMathOperator*{\res}{res}

\newcommand{\nonzero}{\setminus\{0\}}

\newcommand{\de}{\partial}

\newcommand{\half}{\tfrac{1}{2}}

\newcommand{\unpert}{{\rvert}_{\lambda=0}}
\newcommand{\unperte}{{\rvert}_{\epsilon=0}}

\newcommand{\repart}{{\operatorname{\Re\mathfrak{e}}\,}}
\newcommand{\impart}{{\operatorname{\Im\mathfrak{m}}\,}}

\newcommand{\defas}{\mathrel{\raise.095ex\hbox{$:$}\mkern-4.2mu=}}
\newcommand{\defasr}{\mathrel{=\mkern-4.2mu\raise.095ex\hbox{$:$}}}

\newcommand{\mathand}{\quad\text{and}\quad}
\newcommand{\mathwith}{\quad\text{with}\quad}

\newcommand{\beq}{\begin{equation}}
\newcommand{\eeq}{\end{equation}}
\newcommand{\beqn}{\begin{equation*}}
\newcommand{\eeqn}{\end{equation*}}

\DeclareMathAlphabet{\mathfat}{U}{bbold}{m}{n}          
\newcommand{\one}{\mathfat{1}}					

\def\Xint#1{\mathchoice
   {\XXint\displaystyle\textstyle{#1}}%
   {\XXint\textstyle\scriptstyle{#1}}%
   {\XXint\scriptstyle\scriptscriptstyle{#1}}%
   {\XXint\scriptscriptstyle\scriptscriptstyle{#1}}%
   \!\int}
\def\XXint#1#2#3{{\setbox0=\hbox{$#1{#2#3}{\int}$}
     \vcenter{\hbox{$#2#3$}}\kern-.5\wd0}}
\def\regint{\Xint-}

\begin{document}

\title[Asymptotic Expansion of the Homoclinic Splitting Matrix]{Asymptotic Expansion of the Homoclinic Splitting Matrix for the Rapidly, Quasiperiodically, Forced Pendulum}
\author{Mikko Stenlund}
\date{\today}
\email{mikko.stenlund@helsinki.fi}
\address{Department of Mathematics and Statistics, University of Helsinki, P.O. Box 68, Fi-00014 University of Helsinki, Finland}

\begin{abstract}
We study a Hamiltonian describing a pendulum coupled with several ani\-soch\-ro\-nous oscillators, devising an asymptotic expansion for the splitting (matrix) associated with a homoclinic point. This expansion consists of contributions that are manifestly exponentially small in the limit of vanishing hyperbolicity, by a shift-of-contour argument. Hence, we infer a similar upper bound on the splitting itself.

\end{abstract}

\maketitle


\psset{treemode=R,treefit=tight,radius=2pt,tnpos=a,npos=.5,nrot=:U,labelsep=3pt
,treenodesize=-1pt,levelsep=1.2cm}
\newcommand{\rad}{7pt}


\section{Main Concepts and Results}

\subsection{Background and history}
The study of ``separatrix splitting'' dates back to Poincar\'e's classic \textit{Les M\'ethodes Nouvelles de la M\'ecanique C\'eleste} \cite{Poincare3}. 

Starting with Kolmogorov's 1954 note \cite{Kolmogorov}, it was proved in a series of papers over a period of twenty years that quasiperiodic motions (invariant tori) are typical for nearly integrable Hamiltonians \cite{Moser,Arnold2,MoserRapid1,MoserRapid2,Moser67}, and that motions which become quasiperiodic asymptotically in time (stable/unstable manifolds) are stable under small perturbations \cite{Moser67,Graff}. 

Arnold \cite{ArnoldDiffusion} described a mechanism how a chain of such ``whiskered" tori could provide a way of escape for special trajectories, resulting in instability in the system. (A trajectory would typically lie on a torus and therefore stay eternally within a bounded region in phase space.) The latter is often called Arnold mechanism and the general idea of instability goes by the name Arnold diffusion. It is conjectured in \cite{ArnoldAvez} that Arnold diffusion due to Arnold mechanism is present quite generically, for instance in the three body problem.

Arnold mechanism is based on Poincar\'e's concept of biasymptotic solutions, discussed in the last chapter of \cite{Poincare3}, that are formed at intersections of whiskers of tori. Following such intersections, a trajectory can ``diffuse" in a finite time from a neighbourhood of one torus to a neighbourhood of another, and so on. 

Chirikov's report \cite{Chirikov} is a very nice physical account on Arnold diffusion, while Lochak's compendium \cite{LochakCompendium} discusses more recent developments in a readable fashion and is a good point to start learning about diffusion. Gelfreich's introduction \cite{GelfreichLondon} to splitting of separatrices is excellent, and we recommend it to anyone intending to study the topic. From there one should advance to \cite{Gelfreich-review}, which covers more topics with more details. The extensive memoir \cite{LochakMemoirs} by Lochak, Marco, and Sauzin is written from the geometric point of view. It has a historical flavor, making it interesting and accessible to virtually anyone. 

For the separatrix splitting of the \emph{periodically} forced pendulum, see \cite{Holmes,Scheurle,DelshamsPeriodic,EllisonKummerSaenz}. In the case of the standard map, an asymptotic expression of the splitting has been obtained in \cite{GelfreichStandard}, as the culmination of a series of works starting with Lazutkin's \cite{Lazutkin}.

For the \emph{quasiperiodically} forced pendulum, several studies exist. We mention \cite{DelshamsGelfreichJorbaSeara,DelshamsEtal97,DelshamsJorbaSearaGelfreich,DelshamsGutierrez}. Especially \cite{Treschev,LochakMemoirs,RudnevWiggins2,Sauzin,RudnevTen} emphasize geometrical aspects of separatrix splitting in the quasiperiodic setting. A lower bound on the splitting has been obtained in \cite{GallavottiSplit}, where the pendulum is coupled to two rotators---one speeding up and one slowing down. A generalization to the case of several slow rotators can be found in \cite{Procesi}. These results extend \cite{ChierchiaGallavotti,GallavottiTwistless,GentileQuasiflat}. See also \cite{GGMMelnikovDominance}.

\subsection{The model}

We consider the Hamiltonian
\beq
\Ham(\phi,\psi,I,A)=\half I^2+g^2\cos\phi+\half A^2- \lambda f(\phi,\psi)
\label{eq:H}
\eeq
of a pendulum coupled to $d$ rotators, with $\phi\in\es^1\defas\R/2\pi\Z$ and  $I\in\R$ the coordinate and momentum of the pendulum, and $\psi\in\torus\defas(\es^1)^d$ and $A\in\R^d$ the angles and actions of the rotators, respectively. The perturbation $f$ is assumed to be real-valued and real-analytic in its arguments, and $\lambda$ is a (small) real number, whereas $g>0$. This Hamiltonian is sometimes called \emph{the generalized Arnold model} or the \emph{Thirring model}. It is \emph{the} prototype of a nearly integrable Hamiltonian system close to a simple resonance, as is explained in the introduction of \cite{GentileExponent}. A review of applications can be found in \cite{Chirikov}.

The equations of motion are
\beq
\dot\phi=I, \quad \dot\psi=A, \quad 
\dot I=g^2\sin\phi+\lambda\,\de_\phi f,\quad
\dot A=\lambda\,\de_\psi f.
\label{eqm}
\eeq
For the parameter value $\lambda=0$, which is addressed as the unperturbed case, the pendulum and the rotators decouple. The former then has the separatrix flow $\phi:\R\to \es^1$ given by
\beqn
\phi(t)=\Phi^0(e^{gt}) \mathwith 
\Phi^0(z)=4\arctan z.
\eeqn
By elementary trigonometry, the odd function $\Phi^0$ possesses the symmetry property
\beq\label{eq:arctan-sym}
\Phi^0(z)=2\pi-\Phi^0(z\inv).
\eeq
In the phase space of the pendulum, the separatrix---given by $\Phi^0$---separates closed trajectories (libration) from open ones (rotation).



On the other hand, $\psi:\R\to\torus$ is quasiperiodic: the vector $
\omega \defas A(0) \equiv A(t)
$
in
\beqn
\psi(t)=\psi(0)+\omega t\pmod{2\pi}
\eeqn
is assumed to satisfy for some positive numbers $a$ and $\nu$ the Diophantine condition
\beq
|\omega\cdot q| > a\,|q|^{-\nu}\quad{\rm for}\quad q\in \Z^d\nonzero.
\label{Dio}
\eeq
Thus, at the instability point of the pendulum, the flow possesses the invariant tori
\beqn
\T_0\defas\bigl\{(\phi,\psi,I,A)=(0,\theta, 0, \omega)\;\big|\; \theta\in T^d\bigr\}
\eeqn
indexed by $\omega$, with stable and unstable manifolds ($\W^s_0$ and $\W^u_0$, respectively) coinciding:
\beq
\W^{s,u}_0=\bigl\{(\phi,\psi,I,A)=\left(\Phi^0(z), \theta, gz\partial_z\Phi^0(z),\omega\right)\;\big|\; z\in [-\infty,\infty],\;\theta\in T^d\bigr\}.
\label{unpert}
\eeq
\begin{remark}\label{rem:timescale}
The constant $g$ is the Lyapunov exponent for the unstable fixed point of the pendulum motion; in the limit $s\to-\infty$ two nearby initial angles $\phi(s)$ and $\phi(s+\delta s)$ separate at the exponential rate $e^{gs}$. As $\phi(t)=\Phi^0(e^{t/g^{-1}})$, this fixes a natural time scale of $g^{-1}$ units, characteristic of the pendulum motion in the unperturbed Hamiltonian system \eqref{eq:H}.
\end{remark}

When the perturbation is switched on ($\lambda\neq 0$), it is known that most of the invariant tori survive and have stable and unstable manifolds---or ``whiskers'' as Arnold has called them---that may not coincide anymore. We prove a bound on their splitting.

\subsection{Main theorem}
In \cite{Stenlund-whiskers} we construct the perturbed manifolds in a form similar to \eqref{unpert} as graphs of analytic functions over a piece of $[-\infty,\infty]\times\torus$. 
To that end, we look for solutions of the form
\beq
(\phi(t),\psi(t))=(\Phi(e^{\gamma t},\omega t),\omega t+\Psi(e^{\gamma t},\omega t))=(0,\omega t)+(\Phi,\Psi)(e^{\gamma t},\omega t)
\label{trial}
\eeq
with quasiperiodic behavior in \emph{one} of the two limits $t\to \pm\infty$. Note especially that the Lyapunov exponent $\gamma>0$ depends on $\lambda$, with $\gamma\unpert=g$.
\begin{remark}
One should not expect asymptotic quasiperiodicity in both limits $t\to\pm\infty$, as the unstable and stable manifolds, $\W^u_\lambda$ and $\W^s_\lambda$, are generically expected to depart for nonzero values of the perturbation parameter $\lambda$. Therefore, either the past \emph{or} future asymptotic of a trajectory will evolve so as to ultimately reach the deformed invariant torus $\T_\lambda$. 
\end{remark}

Let us denote the total derivative $d/dt$ by $\de_t$ and the complete angular gradient $(\de_\phi,\de_\psi)$ by $\de$ for short. Substituting \eqref{trial} into the equations of motion,
we get the equation
\beqn
(\omega\cdot\de_\theta+\gamma e^{\gamma t}\de_z)^2 X(e^{\gamma t},\omega t)=[(g^2\sin\Phi,0)+\lambda\,\de f(X+(0,\theta))](e^{\gamma t},\omega t)
\eeqn
for $X\defas(\Phi,\Psi)$, where $\theta$ stands for the canonical projection $[-\infty,\infty]\times\torus\to\torus$. 

Notice that the partial differential operator
\beqn
\el\defas\omega\cdot\de_\theta+\gamma z\de_z
\label{el}
\eeqn
satisfies the characteristic identity
\beq\label{Lid}
\el F(ze^{\gamma t},\theta+\omega t)
=\de_t F(ze^{\gamma t},\theta+\omega t),
\eeq
which reflects the time derivative nature of $\el$. In fact, if $T$ is the ``time-reversal map''
\beq\label{eq:time-reversal-map}
T(z,\theta)\equiv (z\inv,-\theta),
\eeq
then, by the chain rule,
\beq\label{eq:time-reversal}
\el (F\circ T)=- (\el F)\circ T.
\eeq

Let us abbreviate
\beq\label{eq:Omega}
\Omega(X) \defas (g^2\sin\Phi,0)+\lambda\,\widetilde\Omega(X)\quad\text{with}\quad
\widetilde\Omega(X)\defas\de f(X+(0,\theta)).
\eeq
We have then encoded the equations of motion into the PDE
\beq\label{xeq}
\el^2 X=\Omega(X).
\eeq

The action variables, or momenta, trivially follow from the knowledge of $X(z,\theta)$:
\beqn
(I(t),A(t))=(0,\omega)+Y(e^{\gamma t},\omega t),\quad Y\defas\el X.
\eeqn
The solutions $X$ provide a parametrization of the deformed tori and their stable and unstable manifolds. 
As hinted below \eqref{trial}, we find two kinds of solutions, $X^u(z,\theta)$ defined for $z\in [-z_0,z_0]$ and $X^s(z,\theta)$ defined for $z\in [-\infty,-z_0\inv]\cup [z_0^{-1},\infty]$. Here, $z_0>1$.

We will consider $\lambda$ small in a $g$-dependent fashion, taking
\beq
\epsilon\defas\lambda g^{-2}
\label{eq:epsilon}
\eeq
small. Such a choice is needed for studying the limit $g\to \infty$, which corresponds to rapid forcing; see Remark~\ref{rem:timescale}. The domain we restrict ourselves to is
\beq\label{eq:D}
D\defas\bigl\{(\epsilon,g)\in \C\times\R\;\big|\;\abs{\epsilon}<\epsilon_0,\; 0<g<g_0\bigr\},
\eeq
for some positive values of $\epsilon_0$ and $g_0$.

The following theorem is from \cite{Stenlund-whiskers}. It is a version of a classical result, and by no means new; earlier treatments include \cite{Melnikov63,Moser67, Graff,EliassonBiasymptotic,GallavottiTwistless,GentileQuasiflat,GentileExponent}. 
\begin{result}[Tori and their whiskers]\label{thm:manifolds}
Let $f$ be real-analytic and even, \ie,
\beqn
f(\phi,\psi)=f(-\phi,-\psi).
\eeqn 
Also, suppose $\omega$ satisfies the Diophantine condition \eqref{Dio}, and fix $g_0>0$. Then there exist a positive number $\epsilon_0$ and a function $\gamma(\epsilon,g)$ on $D$, analytic in $\epsilon$ with $\abs{\gamma-g}<Cg\abs{\epsilon}$, such that equation~\eqref{xeq} has a solution $X^u$ which is analytic in $\epsilon$ as well as in $(z,\theta)$ in a neighbourhood of $\bigunit\times\torus$ and which satisfies ($X^0\defas(\Phi^0,0)$)
\beq
X^u(1,0)=(\pi,0),\quad X^u(z,\theta)=X^0(z)+\order{\epsilon}.
\label{norm}
\eeq
Corresponding to the same $\gamma$, there exists a solution $X^s(z,\theta)=X^0(z)+\order{\epsilon}$ which is an analytic function of $(z^{-1},-\theta)$ in a neighbourhood of $\bigunit\times\torus$. The maps
\beq\label{eq:param}
W^{s,u}(z,\theta)=(X^{s,u},Y^{s,u})(z,\theta)+((0,\theta),(0,\omega)),\quad Y^{s,u}\defas\el X^{s,u},
\eeq
provide analytic parametrizations of the stable and unstable manifolds $\W^{s,u}_\lambda$ of the torus $\mathcal{T}_\lambda$.
\end{result}

\begin{remark}\label{rem:manifolds} 
The number $\epsilon_0$ above depends on the Diophantine exponent $\nu$ and on $f$. The perturbation $(\phi,\psi)\mapsto f(\phi,\psi)$ is analytic on the compact set $\es^1\times\torus$. By Abel's Lemma (multivariate power series converge on polydisks), it extends to an analytic map on a ``strip'' $\abs{\impart\phi},\abs{\impart\psi}\leq \eta$ ($\eta>0$) around $\es^1\times\torus$. By Theorem~\ref{thm:manifolds}, there exists some $0<\sigma<\eta$ such that each $\theta\mapsto X^{s,u}(\piste,\theta)$ is analytic on $\abs{\impart\theta}\leq \sigma$. 
\end{remark}

An important part of Theorem~\ref{thm:manifolds} is that the domains of $X^u$ and $X^s$ overlap. Namely, if $X$ solves equation~\eqref{xeq}, then so does $(2\pi,0)-X\circ T$. This is due to \eqref{eq:time-reversal} and the parity of $f$. Consequently, by time-reversal, the stable and unstable manifolds are related through
\beq\label{Xsym}
X^s=(2\pi,0)-X^u\circ T.
\eeq
In particular, as $T(1,0)=(1,0)$,
$
X^s(1,0)=X^u(1,0).
$
The actions $Y^{s,u}=\el X^{s,u}$ satisfy
$
Y^s=Y^u\circ T,
$
yielding
$
Y^s(1,0)=Y^u(1,0).
$
In other words, a \emph{homoclinic intersection} of the stable and the unstable manifolds $\W^{s,u}_\lambda$ occurs at $(z,\theta)=(1,0)$, as their parametrizations \eqref{eq:param} coincide at this \emph{homoclinic point}. Since the manifolds $\W^{s,u}_\lambda$ are invariant, there in fact exists a \emph{homoclinic trajectory} on which the parametrizations agree:
\beq\label{eq:homoclinic-trajectory}
W^s(e^{\gamma t},\omega t)\equiv W^u(e^{\gamma t},\omega t).
\eeq


Coming to the second result of \cite{Stenlund-whiskers}, let us expand
$
X^u=\sum_{\ell=0}^\infty \epsilon^\ell X^{u,\ell}.
$
It turns out that the common analyticity domain of each $X^{u,\ell}$ in the $z$-variable is in fact much larger than the (small) neighbourhood of $\bigunit$---the corresponding domain of $X^u$ according to Theorem~\ref{thm:manifolds}. Namely, it includes the wedgelike region
\beq\label{eq:treedomain}
\mathbb{U}_{\tau,\vartheta}\defas \bigl\{\abs{z}\leq \tau\bigr\}\,\bigcup\,\bigl\{\arg{z}\in [-\vartheta,\vartheta]\cup [\pi-\vartheta,\pi+\vartheta]\bigr\}\subset\C
\eeq
(with some positive $\tau$ and $\vartheta$). We repeat parts of the argument in Section~\ref{sec:continuation} for convenience. 

\begin{result}[Analytic continuation]\label{thm:extension}
Each order $X^{u,\ell}$ of the solution extends analytically to a common region $\mathbb{U}_{\tau,\vartheta}\times\{\abs{\impart{\theta}}\leq\sigma\}$. Moreover, if $\psi\mapsto f(\piste,\psi)$ is a trigonometric polynomial of degree $N$, \ie, $N$ is the smallest integer such that $\hat f(\piste,q)=0$ whenever $\abs{q}>N$, then $\theta\mapsto X^{u,\ell}(\piste,\theta)$ is a trigonometric polynomial of degree $\ell N$, at most.
\end{result}

\begin{remark}\label{rem:extension}
With $\eta$ and $\sigma$ as in Remark~\ref{rem:manifolds}, the numbers $\tau$ and $\vartheta$ are specified by the following observation: $\Phi^0(z)=4\arctan z$ implies that $\abs{\impart\Phi^0(z)}\leq\eta$ in $\mathbb{U}_{\tau,\vartheta}$ with $\tau$ and $\vartheta$ sufficiently small. By Remark~\ref{rem:manifolds}, $(z,\theta)\mapsto f(\Phi^0(z),\theta)$ is analytic on $\mathbb{U}_{\tau,\vartheta}\times\{\abs{\impart{\theta}}\leq\sigma\}$, which is the basis of the proof.
\end{remark}

In spite of Theorem~\ref{thm:extension}, (a straightforward upper bound on) $X^{u,\ell}$ grows without a limit as $\abs{\repart{z}}\to \infty$, such that there is no reason whatsoever to expect absolute convergence of the series $\sum_{\ell=0}^\infty \epsilon^\ell X^{u,\ell}$ in an unbounded $z$-domain with a fixed $\epsilon$. In fact, it is known that the behavior of the unstable manifold gets extremely complicated for large values of $z$ even with innocent looking Hamiltonian systems.  Still, it seems to us that the possibility of a uniform analytic extension of the coefficients $X^{u,\ell}$ has not been appreciated in the literature.

Due to \eqref{Xsym}, an analog of Theorem~\ref{thm:extension} is seen to hold for  $X^s$, with $z$ replaced by $z\inv$. 

Theorem~\ref{thm:extension} allows one (at each order in $\epsilon$) to track trajectories $t\mapsto W^{s,u}(e^{\gamma t},\theta+\omega t)$ on the invariant manifolds $\W_\lambda^{s,u}$ for arbitrarily long times in a uniform complex neighbourhood $\abs{\impart{t}}\leq g\inv\vartheta$ of the real line, for arbitrary $\theta\in\torus$. The motivation for doing this stems from studying the splitting of the manifolds $\W_\lambda^{s,u}$ in the vicinity of the homoclinic trajectory \eqref{eq:homoclinic-trajectory}. The general ideology that, being able to extend ``splitting related functions" to a large complex domain yields good estimates, is due to Lazutkin \cite{Lazutkin}, as is emphasized in \cite{LochakMemoirs}.

In order to study the intersection more closely, we express the actions as functions of the original angle variables $(\phi,\psi)=X^{s,u}(z,\theta)+(0,\theta)$ appearing in the Hamiltonian \eqref{eq:H}. To this end, let $F^{s,u}:(z,\theta)\mapsto(\phi,\psi)$ be the above coordinate transformations, and write $Y^{s,u}= \bar Y^{s,u}\circ F^{s,u}$. We set
$
\de\defas(\de_\phi,\de_\psi)$ and $D\defas(\de_z,\de_\theta).
$
By the chain rule,
\beq\label{eq:chain}
D Y^{s,u}=\left(\de\bar Y^{s,u}\circ F^{s,u}\right)D F^{s,u}.
\eeq
The invertibility of the matrix $D F^{s,u}$ is a consequence of Theorem~\ref{thm:manifolds}. 
At the homoclinic point $(z,\theta)=(1,0)$, equation~\eqref{eq:homoclinic-trajectory} implies that $F^s(e^{\gamma t},\omega t) \equiv F^u(e^{\gamma t},\omega t)$ and $F^{s,u}(1,0)=(\pi,0)$.

Casting in the obvious manner $\bar Y^{u,s}=(\bar Y^{u,s}_\Phi,\bar Y^{u,s}_\Psi)$, we define the \emph{splitting vector}:
\beqn
\Delta(\phi,\psi)\defas (A^u-A^s)^T=\bigl(\bar Y^{u}_\Psi-\bar Y^{s}_\Psi\bigr)^T(\phi,\psi).
\eeqn
Transposition indicates that we consider $\Delta$ a column vector. We define the \emph{splitting matrices}:
\begin{align}\label{eq:splitmatdef1}
\Upsilon(t) & \defas\de_\theta\!\left(Y^u_\Psi-Y^s_\Psi\right)^T(e^{\gamma t},\omega t), \\\widetilde\Upsilon(t) & \defas \de_\psi\!\left(\bar Y^{u}_\Psi-\bar Y^{s}_\Psi\right)^T(F^{u,s}(e^{\gamma t},\omega t)).
\end{align}
Notice that if $(\phi,\psi)(t)$ stands for the moment for the angles on the homoclinic trajectory at time $t$, with the particular initial condition $(\phi,\psi)(0)=(\pi,0)$, then 
$
\widetilde{\Upsilon}=\de_\psi\Delta(\phi,\psi).
$

It is nontrivial but straightforward to establish that $\widetilde\Upsilon$ and $\Upsilon$ differ by a close-to-identity \emph{factor}, if $\epsilon$ is taken to be small proportionally to $g$:
\begin{result}\label{thm:splitmatrel}
Fix $t\in \R$ and define $\tilde\epsilon\defas g\inv e^{g{\abs{t}}}\epsilon$. Then, as $g\to 0$,
\beqn\label{eq:relation}
\Upsilon(t)=\widetilde\Upsilon(t)\,\bigl(\one+\order{\tilde\epsilon}\bigr),
\eeqn
if $\tilde\epsilon$ is sufficiently small (independently of $g$ and $t$).
\end{result}
Theorem~\ref{thm:splitmatrel} is our first result. Its proof in Appendix~\ref{app:computations} is based on energy conservation and the fact that the actions are given by scalar potentials. We can now state the main theorem:

\begin{result}[Homoclinic splitting]\label{thm:splitting}
Let us assume that $\psi\mapsto f(\piste,\psi)$ is a trigonometric polynomial in addition to the assumptions of Theorem~\ref{thm:manifolds}. Then for each $t\in\R$ there exist positive constants $C$ and $c$, such that the exponentially small upper bound
\beqn
\abs{\Upsilon_{ij}(t)}\leq C\abs{\epsilon} e^{-c g^{-1/(\nu+1)}},
\eeqn
where $\nu$ is the Diophantine exponent, holds, provided $\epsilon g^{-4}$ is sufficiently small.
\end{result}

Theorem~\ref{thm:splitting} is derived from Proposition~\ref{prop:asymptotic}---the centerpiece of this work---stated in Subsection~\ref{subsec:asymptotic}. Technicalities aside, we believe that Subsection~\ref{subsec:asymptotic} can be understood without any further introduction, and we urge the reader to go through it before moving on to Section~\ref{sec:continuation}.

Why just study a $d\times d$ submatrix of the full $(d+1)\times(d+1)$ Jacobi matrix of $(I^u-I^s,A^u-A^s)$ when measuring \emph{transversality} of the homoclinic intersection $\W^s_\lambda\cap\W^u_\lambda$? Because the latter is singular. As a matter of fact, \eqref{eq:identity} simply means that the splitting vanishes in the direction of the homoclinic trajectory. The author is grateful to Dr Mischa Rudnev for pointing this out. For further motivation, see item R1 of Appendix R and Section~8 in \cite{GallavottiSplit}.

\subsection*{Acknowledgements} 
I am indebted to Antti Kupiainen for all his advice. I express my gratitude to Guido Gentile, Kari Astala, and Jean Bricmont for their comments. I thank Giovanni Gallavotti and Emiliano De Simone for discussions at Rutgers University and University of Helsinki, respectively. Mischa Rudnev, Pierre Lochak, and Michela Procesi were kind enough to explain me some of their works on the subject. I am grateful to Joel Lebowitz and Rutgers University for hospitality during the final stages of writing this work. This work was supported by the Finnish Cultural Foundation and NSF Grant DMR-01-279-26.




\section{Analytic Continuation of the Solution}\label{sec:continuation}
Here we present the proof of Theorem~\ref{thm:extension}. We drop the superscript $u$ from the notation and single out the uncoupled part $X^0$ of the complete solution $X$;
\beqn
X=X^0+\widetilde X\mathwith{\widetilde X}\unperte\equiv 0.
\eeqn
Equation~\eqref{xeq} and $\el^2X^0=(\gamma^2\sin\Phi^0,0)$ imply
$
\el^2\widetilde X=-(\gamma^2\sin\Phi^0,0)+\Omega(X^0+\widetilde X).
$
Thus,
\beq\label{til}
\K\widetilde X=\widetilde W(\widetilde X),
\eeq
when we define the linear operator
\beq\label{K}
\K\defas
\begin{pmatrix} 
L & 0\\
0 & \el^2
\end{pmatrix}
\mathwith 
L\defas\el^2-\gamma^2\cos\Phi^0
\eeq
and the nonlinear operator 
\beq\label{tilW}
\widetilde W(\widetilde X)\defas(-\gamma^2\sin\Phi^0-\gamma^2(\cos\Phi^0)\widetilde\Phi,0)+\Omega(X^0+\widetilde X).
\eeq

Throughout the rest of the work, we shall refer to different parts of the Taylor expansion of a suitable function $h(z,\theta)$ around $z=0$ using the notation
\beqn
h_k(\theta)\defas \frac{\de_z^k h(0,\theta)}{k!},\quad h_{\leq k}(z,\theta)\defas \sum_{j=0}^k z^kh_k(\theta)\mathand \delta_k h\defas h-h_{\leq k-1}.
\eeqn

By~\eqref{til}, $\delta_2\widetilde X$ satisfies
\beq\label{Zeq}
\K \delta_2\widetilde X=W(\delta_2\widetilde X),
\eeq
where 
\beq\label{Wdef}
W(Z)\defas\delta_2\left[\widetilde W(\widetilde X_{\leq 1}+Z)+
\begin{pmatrix}
\gamma^2(\cos\Phi^0)\widetilde\Phi_{\leq 1} \\
0
\end{pmatrix}
\right].
\eeq

Let us consider analytic functions $Z$ on a small, compact and complex, neighbourhood $\Pi$ of the set $\bigunit\times\torus$. It is a complex Banach space $\A$, once equipped with the supremum norm, and has the closed subspace
\beq\label{eq:A1}
\A_1\defas\left\{Z\in\A\;|\;Z_{\leq 1}=0\right\}.
\eeq
We have shown in \cite{Stenlund-whiskers} that $\K$ maps $\A_1$ to itself and has a bounded, $\order{\gamma^{-2}}$, inverse.

By virtue of \eqref{Zeq} and $\widetilde W$'s analyticity, $\delta_2\widetilde X$ admits the representation
\beq\label{eq:Zrec}
\begin{split}
\delta_2\widetilde X=\K\inv \delta_2 \biggl[
\begin{pmatrix}
\gamma^2 \cos\Phi^0 & 0 \\
0 & 0
\end{pmatrix}
\widetilde X_{\leq 1}+\sum_{k=0}^\infty w^{(k)}\bigl(\widetilde X_{\leq 1}\bigr)^{\otimes k}
\biggr]+ \\
+\,\K\inv\sum_{k=1}^\infty \Bigl[w^{(k)}\bigl(\widetilde X_{\leq 1}+\delta_2\widetilde X\bigr)^{\otimes k}-w^{(k)}\bigl(\widetilde X_{\leq 1}\bigr)^{\otimes k}\Bigr]
\end{split}
\eeq
on the set $\Pi$, taking $\epsilon$ small enough, and denoting 
\beq\label{eq:wdefinition}
w^{(k)}\defas \frac{1}{k!}D^k\widetilde W(0)
\eeq
as well as a repeated argument of such a symmetric $k$-linear operator by
$
(x)^{\otimes k}\defas  (x,\dots,x),
$
for the sake of brevity. Observe that we have omitted a $\delta_2$ in front of the square brackets on the second line of \eqref{eq:Zrec} as redundant. 

Equation~\eqref{eq:Zrec} may be viewed as a recursion relation for $\delta_2\widetilde X$. It is crucial that 
\beq\label{eq:nodeorder}
w^{(0)},\,w^{(1)}=\order{\epsilon g^2},
\eeq
when $(\epsilon,g)\in D$; see \eqref{eq:D}. Namely, any given order $\delta_2\widetilde X^\ell$ in the convergent expansion
$
\delta_2\widetilde X=\sum_{\ell=1}^\infty \epsilon^\ell \,\delta_2\widetilde X^\ell
$
is then completely determined by $\widetilde X_{\leq 1}$ and the \emph{lower orders} $\delta_2\widetilde X^l$ ($1\leq l\leq\ell-1$) through the right-hand side of \eqref{eq:Zrec}. Moreover, since $\widetilde X_{\leq 1}=\order{\epsilon}$, only \emph{finitely many} terms in the sum over the index $k$ are involved. Together these facts imply that only \emph{finitely many} recursive steps using \eqref{eq:Zrec} are needed to completely describe any given order $\delta_2\widetilde X^\ell$ in terms of $\widetilde X_{\leq 1}$ alone and that, at each such step, only \emph{finitely many} terms from the $k$-sum contribute. 

It is important to understand that $\widetilde X_{\leq 1}$ is a predetermined function. As we shall see, the recursion procedure will then provide the analytic continuation of each $X^{u,\ell}=\widetilde X^\ell_{\leq 1}+\delta_2\widetilde X^\ell$ ($\ell\geq 1$) to the large region $\mathbb{U}_{\tau,\vartheta}\times\{\abs{\impart{\theta}}\leq\sigma\}$ of Theorem~\ref{thm:extension}. 

\subsection{Tree expansion}
We next give a pictorial representation of the above recursion. It involves tree diagrams similar to those of Gallavotti, \textit{et al.} (see, \eg, \cite{GallavottiTwistless,ChierchiaGallavotti}), with one difference: there will be no resummations nor cancellations, as the expansion in \eqref{eq:Zrec} contains no resonances and is instead well converging. This so-called tree expansion is needed for bookkeeping and pedagogical purposes; we simply choose to draw a tree instead of spelling out a formula. 

Let us first define the auxiliary functions
\beqn
h^{(k)}\defas 
\begin{cases}
 w^{(0)}+\bigl[\bigl(\begin{smallmatrix} 
\gamma^2 \cos\Phi^0 & 0 \\
0 & 0
\end{smallmatrix}\bigr)
+ w^{(1)}
\bigr]\widetilde X_{\leq 1} & \text{if $k=1$,} 

\vspace{1mm}\\	

w^{(k)}\bigl(\widetilde X_{\leq 1}\bigr)^{\otimes k} & \text{if $k=2,3,\dots$},
\end{cases}
\eeqn
and make the identifications
\beq\label{eq:endnode}
\raisebox{3pt}{
\pstree{\Tp}{\Tcircle{\tiny $k$}}
}
\defas\K\inv\delta_2 h^{(k)} 
\mathand 
\raisebox{3pt}{
\pstree{\Tp}{\Tc[fillstyle=hlines,fillcolor=black,hatchsep=2.5pt]{\rad}}
}
\defas\K\inv\delta_2 \,\sum_{k=0}^\infty h^{(k)}.
\eeq
Furthermore, let
\beqn
\raisebox{3pt}{
\pstree{\Tp}{\Tc{\rad}}
}
\defas\delta_2\widetilde X,\quad 
\raisebox{3pt}{
\pstree{\Tp}{\TC*}
} 
\defas \widetilde X_{\leq 1}, \mathand 
\raisebox{3pt}{
\pstree[treesep=1.2cm]{\Tp}{\pstree{\TC*~[tnpos=r,tnsep=7.5pt]{\raisebox{1pt}{\tiny $k$ lines}}}{\Tp[name=top] \Tp[name=bot]}}
\ncarc[linestyle=dotted,nodesep=5pt]{top}{bot}  
}
\defas \K\inv w^{(k)}.
\vspace{1mm}
\eeqn
In the diagram representing the $k$-linear $w^{(k)}$, the $k$ ``free'' lines to the right of the node stand for the arguments. We say that these lines \emph{enter} the \emph{internal} node, whereas the single line to the left of the node \emph{leaves} it. For instance,
\beqn
\raisebox{3pt}{
\pstree[treesep=0.5cm]{\Tp}{\pstree{\TC*}{\TC* \Tc[bbh=12pt]{\rad} \Tcircle[]{\tiny 4}}}
}
=\K\inv w^{(3)}\bigl(\widetilde X_{\leq 1},\,\delta_2\widetilde X,\,\K\inv\delta_2 h^{(4)}\bigr).
\eeqn
Notice that, as $w^{(k)}$ is symmetric, permuting the lines entering a node does not change the resulting function. We emphasize that all of the functions introduced above are analytic on $\Pi$ and $\abs{\epsilon}<\epsilon_0$. 

In terms of such \emph{tree diagrams}, or simply \emph{trees}, equation~\eqref{eq:Zrec} reads
\beq\label{eq:tree-rec}
\begin{split}
\raisebox{3pt}{
\pstree{\Tp}{\Tc{\rad}}
}
= \, &
\raisebox{3pt}{
\pstree{\Tp}{\Tc[fillstyle=hlines,fillcolor=black,hatchsep=2.5pt]{\rad}}
}
+
\raisebox{3pt}{
\pstree{\Tp}{\pstree{\TC*}{\Tc{\rad}}}
}
+
\raisebox{3pt}{
\pstree{\Tp}{\pstree{\TC*}{\Tc{\rad} \TC*}}
}
+
\\
& +
\raisebox{3pt}{
\pstree{\Tp}{\pstree{\TC*}{\TC* \Tc{\rad}}}
}
+
\raisebox{3pt}{
\pstree{\Tp}{\pstree{\TC*}{\Tc{\rad} \Tc{\rad}}}
}
+\,\cdots,
\end{split}
\eeq
using multilinearity to split the sums $\widetilde X_{\leq 1}+\delta_2\widetilde X$ into pieces. Above, the sum after the first tree consists of \emph{all} trees having one internal node and an arbitrary number of \emph{end nodes, at least one of which, however, is a white circle}. This rule encodes the fact that on the second line of \eqref{eq:Zrec} the summation starts from $k=1$ and that the contributions with only $\widetilde X_{\leq 1}$ in the argument (\ie, trees with only black dots as end nodes) are cancelled.

Using \eqref{eq:Zrec} recursively now amounts to replacing each of the lines with a white-circled end node by the complete expansion of such a tree above. This is to be understood additively, so that replacing one end node, together with the line leaving it, by a sum of two trees results in a sum of two new trees. For example, such a replacement in the third tree on the right-hand side of \eqref{eq:tree-rec} by the first two trees gives the sum
\beqn
\raisebox{3pt}{
\pstree{\Tp}{\pstree{\TC*}{\Tc[fillstyle=hlines,fillcolor=black,hatchsep=2.5pt]{\rad} \TC*}}
}
+
\raisebox{3pt}{
\pstree{\Tp}{\pstree{\TC*}{\pstree{\TC*}{\Tc{\rad}}  
\TC*}}
}.
\eeqn

Before proceeding, we introduce a little bit of terminology. The leftmost line in a tree is called the \emph{root line}, whereas the node it leaves (\ie, the uniquely defined leftmost node) is called the \emph{root}. A line leaving a node $v$ and entering a node $v'$ can always be interpreted as the root line of a \emph{subtree}, the maximal tree consisting of lines and nodes in the original tree with $v$ as its root. We call $v$ a (not necessarily unique) \emph{successor} of $v'$, whereas $v'$ is the unique \emph{predecessor} of $v$.

The recursion \eqref{eq:tree-rec} can be repeated on a given tree if it has at least one white circle left. Otherwise, the tree in question must satisfy 
\begin{itemize}
\item[{\bf (R$\mathbf 1'$)}] The tree has only filled circles (\raisebox{3pt}{\Tc[fillstyle=hlines,fillcolor=black,hatchsep=2.5pt]{\rad}}) and black dots (\raisebox{3pt}{\TC*}) as its end nodes,
\end{itemize}
together with
\begin{itemize}
\item[{\bf (R$\mathbf 2'$)}] Any internal node has an entering (line that is the root line of a) subtree containing at least one filled circle as an end node.
\end{itemize}
After all, the recursion can only stop by replacing an existing white circle with a filled one. Continuing \textit{ad infinitum} yields the expansion
\beq\label{eq:tree-exp1}
\raisebox{3pt}{
\pstree{\Tp}{\Tc{\rad}}
}
=\,\sum{\bigl(\text{Trees satisfying (R$1'$) and (R$2'$)}\bigr)}=\, \sideset{}{'}\sum_{\text{trees $T$}}T,
\eeq
where the prime restricts the summation to trees $T$ satisfying (R$1'$) and (R$2'$). We point out that each admissible tree appears precisely once in this sum, considering different two trees that can be superposed by a (nontrivial) permutation of subtrees that enter the same node.

The earlier discussion concerning the description of $\delta_2\widetilde X^\ell$ in terms of a finite sum involving only $\widetilde X_{\leq 1}$ translates to the language of trees in a straightforward fashion. First, the second part of \eqref{eq:nodeorder} and $\widetilde X_{\leq 1}=\order{\epsilon}$ amount pictorially to
\vspace{1mm}
\beqn
\raisebox{3pt}{
\pstree{\Tp}{\pstree{\TC*}{\Tp}}
}
=\order{\epsilon}
\mathand
\raisebox{3pt}{
\pstree{\Tp}{\TC*}
} 
=\order{\epsilon},
\vspace{1mm}
\eeqn
since $\K\inv$ produces a factor of $g^{-2}$. Second, $w^{(k)}=\order{g^2}$ and the first part of \eqref{eq:nodeorder} yield
\beqn
\raisebox{3pt}{
\pstree{\Tp}{\Tcircle{\tiny $k$}}
} 
=\order{\epsilon^k}\quad(k\geq 1)
\qquad \text{and} \qquad
\raisebox{3pt}{
\pstree[treesep=1.2cm]{\Tp}{\pstree{\TC*~[tnpos=r,tnsep=7.5pt]{\raisebox{1pt}{\tiny $k$ lines}}}{\Tp[name=top] \Tp[name=bot]}}
\ncarc[linestyle=dotted,nodesep=5pt]{top}{bot}  
}
=\order{1}\quad(k\geq 2).
\vspace{2mm}
\eeqn
Expanding the filled end nodes
\beq\label{eq:node-exp}
\raisebox{3pt}{
\pstree{\Tp}{\Tc[fillstyle=hlines,fillcolor=black,hatchsep=2.5pt]{\rad}}
}
=\,\sum_{k=1}^\infty 
\raisebox{3pt}{
\pstree{\Tp}{\Tcircle{\tiny $k$}}
},
\eeq
according to \eqref{eq:endnode}, on the right-hand side of \eqref{eq:tree-exp1}, we get a new version of the latter by replacing the rules (R$1'$) and (R$2'$), respectively,  with
\begin{itemize}
\item[{\bf (R1)}] The tree has only numbered circles (\raisebox{3pt}{{\tiny \Tcircle{$k$}}} with arbitrary values of $k$) and black dots (\raisebox{3pt}{\TC*}) as its end nodes,
\end{itemize}
and
\begin{itemize}
\item[{\bf (R2)}] Any internal node has an entering (line that is the root line of a) subtree containing at least one numbered circle as an end node.
\end{itemize}
Let us define the \emph{degree} of a tree as the positive integer
\beq\label{eq:deg}
\deg{T}\defas\#(\!
\raisebox{3pt}{
\pstree{\Tp}{\pstree{\TC*}{\Tp}}
} 
\!)\,+\,\#(\!
\raisebox{3pt}{
\pstree{\Tp}{\TC*}
}
\!)\,+\,\sum_{k=1}^\infty \,k\,\#(\!
\raisebox{3pt}{
\pstree{\Tp}{\Tcircle{\tiny $k$}}
}
\!)
\eeq
for any tree $T$ satisfying (R1) and (R2). By $\#(G)$ we mean the number of occurrences of the graph $G$ in the tree $T$. That is, the degree of a tree is the number of its end nodes with suitable weights plus the number of nodes with precisely one entering line. Since a tree has finitely many nodes, its degree is well-defined. Then a rearrangement of the sum arising from \eqref{eq:node-exp} being inserted into \eqref{eq:tree-exp1} yields formally
\beq\label{eq:tree-exp2}
\raisebox{3pt}{
\pstree{\Tp}{\Tc{\rad}}
}
=\,\sum_{l=1}^\infty\, \sideset{}{^*}\sum_{\substack{\text{trees $T$} \vspace{1pt} \\ \deg{T}=l}}T,
\eeq
where the asterisk reminds us that the rules (R1) and (R2) are being respected.

According to the analysis above, the particular graphs appearing in the definition of $\deg{T}$ are the only possible single-node subgraphs of $T$ proportional to a positive power of $\epsilon$. Since each tree is analytic in $\epsilon$, writing $(\piste)^k$ for the $k$th coefficient of the power series in $\epsilon$, we have
\beqn
T=\sum_{k=\deg T}^\infty \epsilon^k \,T^k=\epsilon^{\deg T}\,\sum_{k=0}^\infty \epsilon^k \,T^{k+\deg T}.
\eeqn
Hence, only trees with degree \emph{at most} equal to $\ell$ can contribute to $\delta_2\widetilde X^\ell$: 
\beq\label{eq:tree-exp-order}
\delta_2\widetilde X^\ell=\,\sum_{l=1}^\ell\, \sideset{}{^*}\sum_{\deg{T}=l}T^\ell=\, {\Bigl(\;
\sideset{}{^*}\sum_{\deg{T}\leq \ell}T
\Bigr)}^{\ell}
\eeq
 or, alternatively,
\beq\label{eq:tree-exp3}
\delta_2\widetilde X=\, \sideset{}{^*}\sum_{\deg{T}\leq \ell}T+\order{\epsilon^{\ell+1}}\qquad(\epsilon\to 0)
\eeq
for \emph{each and every} $\ell=1,2,\dots$. The expansion in \eqref{eq:tree-exp2} is just a compact way of writing \eqref{eq:tree-exp3}. We emphasize that the latter can be derived completely rigorously, for each value of $\ell$ separately, but resorting to the use of formal series allowed us to treat all orders of $\delta_2\widetilde X$ at once. We call the series \eqref{eq:tree-exp2} an \emph{asymptotic expansion} of $\delta_2\widetilde X$; the partial sums ${\sum_{\deg{T}\leq \ell}^*}\,T$ need not converge to $\delta_2\widetilde X$ for any fixed $\epsilon$ as $\ell\to\infty$, but for a fixed $\ell$ the error is bounded by an $\ell$-dependent constant times $\abs{\epsilon}^{\ell+1}$ on the mutual domain of analyticity, $\abs{\epsilon}<\epsilon_0$.
\begin{example}
The beginning of the asymptotic expansion \eqref{eq:tree-exp3} reads
\vspace{1mm}
\beqn
\psset{levelsep=1cm}
\begin{split}
\delta_2\widetilde X
\quad = \quad \, & \raisebox{3pt}{
\pstree{\Tp}{\Tcircle{\tiny 1}}
}
+\order{\epsilon^2} \quad = \quad
\raisebox{3pt}{
\pstree{\Tp}{\Tcircle{\tiny 1}}
}
+
\raisebox{3pt}{
\pstree{\Tp}{\Tcircle{\tiny 2}}
}
+
\raisebox{3pt}{
\pstree{\Tp}{\pstree{\TC*}{\Tcircle{\tiny 1}}}
}
+  \\[2mm] 
& \, +
\raisebox{3pt}{
\pstree{\Tp}{\pstree{\TC*}{\Tcircle{\tiny 1} \TC*}}
}
+
\raisebox{3pt}{
\pstree{\Tp}{\pstree{\TC*}{\TC* \Tcircle{\tiny 1}}}
}
+
\raisebox{3pt}{
\pstree{\Tp}{\pstree{\TC*}{\Tcircle{\tiny 1} \Tcircle{\tiny 1}}}
}
+\order{\epsilon^3}.
\end{split}
\eeqn
\end{example}

\subsection{Analyticity domain of trees} 
As already pointed out, all trees $T$ above are analytic functions of $(z,\theta,\epsilon)$ on $\Pi\times\{\abs{\epsilon}<\epsilon_0\}$. Due to the projections $\delta_2$ appearing in \eqref{eq:endnode}, they also satisfy $T|_{z=0}=\de_zT|_{z=0}=0$, \ie, are elements of the space $\A_1$ defined in \eqref{eq:A1}. On this space, the operator $\K$ in \eqref{K} has a bounded inverse given by a simple integral kernel \cite{Stenlund-whiskers}. Consequently, the analyticity domain of a tree in the $z$-variable is in fact much larger than the neighbourhood of $[-1,1]$ included in $\Pi$; it contains the wedgelike region $\mathbb{U}_{\tau,\vartheta}$ in \eqref{eq:treedomain}.

\begin{lemma}[Analytic continuation of trees]\label{lem:an-cont}
Without affecting the analyticity domain with respect to $\epsilon$, there exist numbers $0<\tau<1$, $0<\vartheta<\pi/2$, and $0<\sigma<\eta$ such that each tree in the sums \eqref{eq:tree-exp2} and \eqref{eq:tree-exp3} extends to an analytic function of $(z,\theta)$ on $\mathbb{U}_{\tau,\vartheta}\times\{\abs{\impart{\theta}}\leq\sigma\}$.
\end{lemma}


Since the sum in \eqref{eq:tree-exp-order} is finite and the functions $\widetilde X_{\leq 1}$ and $X^0$ in $X=X^0+\widetilde X_{\leq 1}+\delta_2\widetilde X$ are analytic on $\mathbb{U}_{\tau,\vartheta}\times\{\abs{\impart{\theta}}\leq\sigma\}$, Theorem~\ref{thm:extension} follows from Lemma~\ref{lem:an-cont}.

\section{Size of the Homoclinic Splitting}

The Lyapunov exponent $\gamma=\gamma(\epsilon,g)$ is an analytic function of $\epsilon$ in a neighbourhood of the origin ($\abs{\epsilon}<\epsilon_0$). In fact,
$
\gamma=g(1+\order{\epsilon})
$
for small $\epsilon$. Further, since $\el$ depends on $\gamma$, we have 
$
\el=\el_0+\order{\epsilon}
$
with
$
\el_0= \omega\cdot\de_\theta+gz\de_z
$
. In particular, as $X^0(z,\theta)\equiv(4\arctan z,0)^T$,
the expansion of the splitting matrix $\Upsilon$---see \eqref{eq:splitmatdef1}---begins linearly with respect to $\epsilon$:
\beqn
\Upsilon=\sum_{\ell=1}^\infty\epsilon^\ell\Upsilon^\ell.
\eeqn
It is customary to call the first order coefficient, $\Upsilon^1$, the \emph{Melnikov term} or the \emph{Melnikov matrix}.

\subsection{The Melnikov term}
Let us study $\Upsilon^1$. The symmetry \eqref{eq:arctan-sym} comes in very handy, since it implies, together with the even parity of $f$, that
$
f(\Phi^0(z\inv),-\theta)\equiv f(\Phi^0(z),\theta).
$
Setting
\beq\label{eq:melnikov}
F(s,\theta)\equiv\int_{-\infty}^\infty \de_\psi^2 f(\Phi^0(e^{g t+s}),\theta+\omega t)-\de_\psi^2 f(0,\theta+\omega t)\,dt
\eeq
then yields, after some work,
\beqn
\Upsilon^1(t)= g^2 F(0,0)
\eeqn
for all $t$. Here the second term in the integrand removes the quasiperiodic limit of the first one, making the integral absolutely convergent. In fact, the integrand tends to zero at an exponential rate as $\abs{t}\to\infty$.

We now give a result, due to Lazutkin in it's original form, \cite{Lazutkin}, and extended by \cite{DelshamsGelfreichJorbaSeara}, \cite{Sauzin}, and \cite{LochakMemoirs} to the quasiperiodic setting. See Appendix~\ref{app:computations} for the proof.
\begin{lemma}[Lazutkin]\label{lem:Lazutkin}
Suppose a function $F$ is analytic on $S\defas[-i\vartheta,i\vartheta]\times\{\abs{\impart \theta}<\eta\}$, continuous on the closure $\bar S$, and satisfies the identity
\beq\label{eq:Lazutkin-id}
F(s,\theta)\equiv F(0,\theta-g\inv\omega s).
\eeq
Then $F$ extends analytically to $\{\abs{\impart s}\leq\vartheta\}\times \{\abs{\impart \theta}<\eta\}$, with \eqref{eq:Lazutkin-id} still holding. On $\R\times\torus$ it obeys a bound of the form
\beqn
\abs{F(s,\theta)-\widetilde F}\leq C B(g) e^{-c\,g^{-1/(\nu+1)}},\quad B(g)\defas \sup_{(s,\theta)\in\bar S}{\abs{F(s,\theta)}},
\eeqn
where $\widetilde F$ stands for the $\theta$-average $\average{F(s,\piste)}$ and is independent of $s$.
\end{lemma}

\begin{remark}
Notice that, for small values of $g$, the analyticity domain of $F$ in Lemma~\ref{lem:Lazutkin} is much larger than what the right-hand side of \eqref{eq:Lazutkin-id} suggests.
\end{remark}

The lemma applies to the function defined in \eqref{eq:melnikov}, due to Remarks~\ref{rem:manifolds} and \ref{rem:extension} regarding the analyticity of $f$. Because the $\psi$-derivatives in the integrand are interchangeable with total $\theta$-derivatives, $\widetilde F=0$, and we get
\begin{proposition}\label{prop:melnikov}
There exist positive constants $c$ and $C$ such that the Melnikov term satisfies
\beqn
\abs{\Upsilon^1}\leq C g^2 e^{-c\,g^{-1/(\nu+1)}}.
\eeqn
\end{proposition}

\begin{remark}
Identity~\eqref{eq:Lazutkin-id} in Lazutkin's lemma is equivalent to
\beq\label{eq:LazutkinPDE}
(\de_s+g\inv\omega\cdot\de_\theta)F=0,
\eeq
or, calling $\bar F(z,\theta)\equiv F(g\inv\ln z,\theta)$, to $\el_0\bar F=0$. There exists a whole industry trying to push through the argument given above replacing $\Upsilon^1$ with a ``better measure'' of the splitting, such as $\Upsilon$, by searching for a coordinate system in which the ``measure'' satisfies \eqref{eq:LazutkinPDE}. The state of this quest is best described in \cite{LochakMemoirs}.
\end{remark}

Neither Lemma~\ref{lem:Lazutkin} nor Proposition~\ref{prop:melnikov} is optimal. There is a refinement in \cite{Sauzin} which, however, is not optimal. In many special cases one can derive a much stronger bound.

\subsection{Regularized integrals}
This subsection borrows heavily from Gallavotti,~\textit{et al.}; see for instance \cite{ChierchiaGallavotti} and \cite{GallavottiSplit}. We adhere to the notation
\beqn
X(t;z,\theta)\defas X(ze^{\gamma t},\theta+\omega t),
\eeqn
which makes the dependence on the Lyapunov exponent $\gamma$ \emph{implicit} (Recall that $\gamma$ is a function of $\epsilon$), such that the perturbation expansion reads compactly
$
X(t;z,\theta)=\sum_{\ell=0}^\infty \epsilon^\ell X^\ell(t;z,\theta).
$
With little trickery, the equations of motion retain their appearance: \emph{defining}
$
\Omega(X)(t;z,\theta)  \equiv [(g^2\sin\Phi,0)+\lambda\,\widetilde\Omega(X)](t;z,\theta)
$
together with
$
\widetilde\Omega(X)(t;z,\theta)  \equiv\de f(X(t;z,\theta)+(0,\theta+\omega t)),
$
and implementing \eqref{Lid}, we obtain the analogue of \eqref{xeq}:
\beq\label{eq:eqm}
\de_t Y(t;z,\theta)\defas\de_t^2 X(t;z,\theta)=\Omega(X)(t;z,\theta).
\eeq

There is a need to consider functions $h(t;z,\theta)$ that can be expanded as finite sums
\beq\label{eq:t-expansion}
\sum_{p=0}^P t^p h^p(z e^{\gamma t},\theta+\omega t), 
\eeq
where $h^p$ is analytic on $\{0<\abs{z}<\tau\}\times\torus$---a punctured neighbourhood of the origin times the $d$-torus---with at worst a \emph{finite} pole at $z=0$.

Let us set $T_N h(t;z,\theta)\defas \sum_{\abs{q}\leq N}\hat h(t;z,q)e^{iq\cdot\theta}$ and define a \emph{regularized integral}
\beqn
\regint h\,(t;z,\theta)\defas \lim_{N\to\infty}\underset{R=0}{\textrm{res}}\,\frac{1}{R}\int_{-\infty}^t e^{-R\abs{\tau}} T_N h(\tau;z,\theta)\,d\tau,
\eeqn
the residue at $R=0$ meaning that of the analytic extension from the complex half-plane with $\repart{R}$ sufficiently large. This is well-known from the theory of Laplace transforms. The truncation, $T_N$, is needed to insure that the origin of the $R$-plane is not an accumulation point for poles, since factors $(R+i\omega\cdot q)\inv$ appear from the integral.

By the following proposition, it is natural to employ the compelling notation
\beqn
\regint_{-\infty}^t h(\tau;z,\theta)\,d\tau\defas \regint h\,(t;z,\theta),
\eeqn
and to view $z$ and $\theta$ as mere parameters: 
\begin{proposition}\label{prop:regint}
The regularized integral is linear. Moreover,
\begin{enumerate}
\item $\regint h\,(t;z,\theta)=\regint h\,(t_0;z,\theta)+\int_{t_0}^t h(\tau;z,\theta)\,d\tau$ for any real $t_0$, 
\item if $h$ and $h'$ are trigonometric polynomials (in $\theta$) and $h(t;z,\theta)=h'(t;z,\theta)$ for all $t$, then $\regint h\,(t;z,\theta)=\regint h'\,(t;z,\theta)$ for all $t$,
\item $T_N\,\regint h=\regint T_Nh$,
\item $
\de_t\,\regint h\,(t;z,\theta)=\frac{d}{dt}\,\regint_{-\infty}^t h(\tau;z,\theta)\,d\tau=h(t;z,\theta),
$ 
\item $
\regint\!\!\regint h\,(t;z,\theta) \defasr \regint_{-\infty}^t\regint_{-\infty}^\tau h(s;z,\theta)\,ds\,d\tau=\regint_{-\infty}^t (t-\tau)\,h(\tau;z,\theta)\,d\tau.
$
\end{enumerate}
\end{proposition}
\begin{proof}
For a finite $t_0$, $\textrm{res}_{R=0} \frac{1}{R}\int_{t_0}^t e^{-R\abs{\tau}} u(\tau)\,d\tau$ equals the evaluation $\int_{t_0}^t e^{-R\abs{\tau}} u(\tau)\,d\tau\big|_{R=0}=\int_{t_0}^t u(\tau)\,d\tau$, since the integral is entire in $R$, from which (1) follows. If $h$ is a trigonometric polynomial, then $\regint h\,(t;z,\theta)=\underset{R=0}{\textrm{res}}\,\frac{1}{R}\int_{-\infty}^t e^{-R\abs{\tau}} h(\tau;z,\theta)\,d\tau$, yielding (2). Claim (3) is trivial and (4) follows from (1). In order to prove (5), one first checks that it holds for trigonometric polynomials and then deals with the general case.
\end{proof}
We also define
\beqn
\regint_t^\infty h(\tau;z,\theta)\,d\tau \defas \lim_{N\to\infty}\underset{R=0}{\textrm{res}}\,\frac{1}{R}\int_t^{\infty} e^{-R\abs{\tau}} T_N h(\tau;z,\theta)\,d\tau
\eeqn
when it makes sense, 
and write
\beqn
\regint_{\pm\infty}^t=-\,\regint_t^{\pm\infty}\mathand  \regint_{-\infty}^t+\,\,\regint_t^\infty=\regint_{-\infty}^\infty.
\eeqn

The operator $\D\inv$ is defined by giving the expression of its Fourier kernel: 
\beqn
\D\inv (p,q)=
\begin{cases}
(i\omega\cdot q)^{-1} & \text{if $p=q\in\Z^d\nonzero$}, \\
0 & \text{otherwise}.
\end{cases} 
\eeqn
It is the formal inverse of $\D$ on functions $h(z,\theta)$ with vanishing $\theta$-average. Finally,
\beqn
\I h(z,\theta)\defas\int_{-\infty}^0 h(ze^{\gamma\tau},\theta+\omega\tau)\,d\tau.
\eeqn
Applying \eqref{Lid} in the case $F=\I h$, the formal identity
$
\el\I=\one
$
follows. Recalling $\delta_1 h(z,\theta)\defas h(z,\theta)-h_0(\theta)=h(z,\theta)-h(0,\theta)$, one also gets
$
\I\el\delta_1 =\delta_1
$
by a direct computation.

\begin{proposition}\label{prop:regint&I}
Let $h$ be analytic on $\{\abs{z}<\tau\}\times\torus$, and denote $h(t;z,\theta)\equiv h(ze^{\gamma t},\theta+\omega t)$. 
\beqn
\regint_{-\infty}^t h(\tau;z,\theta)\,d\tau=\average{h_0}t+\D\inv h_0(\theta+\omega t)+\I\delta_1 h(ze^{\gamma t},\theta+\omega t)
\eeqn
and
\beqn
\regint_{-\infty}^t \de_\tau h(\tau;z,\theta)\,d\tau=h(t;z,\theta)-\average{h_0}.
\eeqn
\end{proposition}
\begin{proof}
Decompose $h=\average{h_0}+[h_0-\average{h_0}]+\delta_1 h$ and use the linearity of $\regint$. For the second identity, notice $\de_\tau h(\tau;z,\theta)=\D h_0(\theta+\omega \tau)+\el\delta_1 h(ze^{\gamma \tau},\theta+\omega \tau)$, and recall the identities $\D\inv\D h_0=h_0-\average{h_0}$ and $\I\el\delta_1 h=\delta_1 h$.
\end{proof}

\begin{remark}
Observe that, unless the integrals converge in the traditional sense, $\regint_{-\infty}^t$ does generically \emph{not} tend to $\regint_{-\infty}^\infty$ or 0 as $t$ tends to $\infty$ or $-\infty$, respectively.
\end{remark}

A detailed proof of the following elementary result can be found in \cite{Thesis}.
\begin{proposition}\label{prop:Kinverse}
Let $\mathcal{S}_0$ and $\mathcal{S}_1$ be the spaces of analytic functions $h$ on $\{\abs{z}<\tau\}\times\torus$ satisfying $\average{h_0}=0$ and  $\average{h_1}=0$, respectively. Denote $h(t;z,\theta)\equiv h(ze^{\gamma t},\theta+\omega t)$. The operators $L_t=\de_t^2-\gamma^2\cos\Phi^0(z e^{\gamma t})$, $L=\el^2-\gamma^2\cos\Phi^0$, and $\el=\D+\gamma z\de_z$ satisfy 
\beqn
\de_t h(t;z,\theta)\equiv \el h(ze^{\gamma t},\theta+\omega t) \mathand L_th(t;z,\theta)\equiv Lh(ze^{\gamma t},\theta+\omega t).
\eeqn
The operator $\regint$ maps $\mathcal{S}_0$ into itself and, in fact,
$\regint=\de_t\inv$ on $\mathcal{S}_0$.
Similarly, the operator defined by $\regint_{-\infty}^t K_\Phi(t,\tau;z)\,h(\tau;z,\theta)\,d\tau$ maps $\mathcal{S}_1$ into itself and is the inverse of $L_t$ on $\mathcal{S}_1$.
In particular, the operator $\K=\bigl(\begin{smallmatrix} L & 0 \\ 0 & \el^2 \end{smallmatrix}\bigr)=\bigl(\begin{smallmatrix} L_t & 0 \\ 0 & \de_t^2 \end{smallmatrix}\bigr)$ is invertible on the space $\mathcal{S}_1\times\mathcal{S}_0$, where
\beqn
\K\inv h(t;z,\theta)\equiv \regint_{-\infty}^t K(t,\tau;z)h(\tau;z,\theta)\,d\tau. 
\eeqn
\end{proposition}
Above, the kernel $K=\bigl(\begin{smallmatrix} K_\Phi & 0 \\ 0 & K_\Psi \end{smallmatrix}\bigr)$ splits into the useful sum
\beq\label{eq:Ksplit}
K(t,\tau;z)=\sum_{i=0}^1 \sum_{j=1}^2 (t-\tau)^i K_{ij}(t;z)\bar K_{ij}(\tau;z),
\eeq
setting $P(t;z) \defas (z^2 e^{2\gamma t}+1)\inv ze^{\gamma t}$, $Q(t;z) \defas z\inv e^{-\gamma t} (z^2 e^{2\gamma t}-1)$, and
\begin{xalignat*}{3}
K_{01} &=\frac{1}{2\gamma}
\begin{pmatrix}
Q & 0 \\
0 & 0
\end{pmatrix},
&
\bar K_{01} &= \bar K_{11} =
\begin{pmatrix}
P & 0 \\
0 & 0
\end{pmatrix},
&
K_{02} &=-\frac{1}{2\gamma}
\begin{pmatrix}
P & 0 \\
0 & 0
\end{pmatrix},
\\
\bar K_{02} &=
\begin{pmatrix}
Q & 0 \\
0 & 0
\end{pmatrix},
&
K_{11} &=2
\begin{pmatrix}
P & 0 \\
0 & 0
\end{pmatrix},
&
K_{12}&=\bar K_{12}=
\begin{pmatrix}
0 & 0 \\
0 & \one
\end{pmatrix}.
\end{xalignat*}

\subsection{Asymptotic expansion for the splitting matrix} \label{subsec:asymptotic}
Starting from \eqref{eq:eqm} and using Proposition~\ref{prop:Kinverse} with $\average{Y_0^u}=\average{\D X_0^u}=0$, we have
\beqn
Y^u(t;z,\theta)=\regint_{-\infty}^t\Omega(X^u)(\tau;z,\theta) \,d\tau
\eeqn
The identity $X^s(t;z,\theta)\equiv (2\pi,0)-X^u(-t;z\inv,-\theta)$ yields
$
Y^s(t;z,\theta)\equiv Y^u(-t;z\inv,-\theta),
$
such that defining the $d$ component column vector
$
\Delta(t;z,\theta)\defas (Y_\Psi^u-Y_\Psi^s)(t;z,\theta)
$
and the functions 
$
f^{s,u}(t;z,\theta)\defas f(X^{s,u}(t;z,\theta)+(0,\theta+\omega t))
$
thus results in an expression for the splitting matrix $\Upsilon$ of \eqref{eq:splitmatdef1} in terms of regularized integrals: we get
\beq\label{eq:splitting-homoclinic}
\Upsilon(t)=\de_\theta\Delta(t;1,0)=\de_\theta\Delta(0;e^{\gamma t},\omega t)
\eeq
because
\beqn
\de_\theta\Delta(t;z,\theta)=\lambda\,\regint_{-\infty}^t \de_\theta f^u_\psi(\tau;z,\theta)\,d\tau +\lambda\,\regint_{t}^\infty \de_\theta f^s_\psi(\tau;z,\theta)\,d\tau,
\eeqn
where the subindices attached to $f^{s,u}$ stand for partial derivatives (our convention is $\varphi=(\phi,\psi)$ and $\de=\de_{\varphi}=(\de_\phi,\de_\psi)$). 
\begin{remark}
Extracting here the first order in $\epsilon=g^{-2}\lambda$ casts the Melnikov matrix in a compact form in terms of the regularized integrals:
\beqn
\Upsilon^1= g^2\,\regint_{-\infty}^\infty \de_\psi^2 f(\Phi^0(e^{g\tau}),\omega \tau)\,d\tau.
\eeqn
The reader is invited to compare this with \eqref{eq:melnikov} and the equation following it. We point out that a similar procedure, using Lemma~\ref{lem:Lazutkin}, that resulted in the exponentially small bound on $\Upsilon^1$ can be applied here, despite the unusual integrals.
\end{remark}
Suppose that, even with the superscript $u$, the integral 
\beq\label{eq:problem-integral}
\regint_{t}^\infty \de_\theta f^u_\psi(\tau;z,\theta)\,d\tau
\eeq
exists. If the integrand is a trigonometric polynomial in $\theta$, then Proposition~\ref{prop:regint} implies that the value of the integral only depends on the integrand's restriction to $\{(t;z,\theta)\,|\,t\in\R\}$, as the notation suggests \footnote{This is the first of the two places where we need to restrict ourselves to trigonometric polynomials\ldots}. Now, \emph{fixing} $(z,\theta)=(1,0)$ and dropping it from the notation,
\beq\label{eq:splitting}
\Upsilon(t)=\lambda\,\regint_{-\infty}^\infty \de_\theta f^u_\psi(\tau)\,d\tau +\lambda\,\regint_{t}^\infty \bigl[f^s_{\psi\varphi}\,\de_\theta(X^s-X^u)\bigr](\tau)\,d\tau,
\eeq
since $X^u(\tau;1,0)\equiv X^s(\tau;1,0)$. 

Let us make the assumption that $\psi\mapsto f(\piste,\psi)$ is a trigonometric polynomial, such that, by Theorem~\ref{thm:extension}, each order of $X$ in $\epsilon$ is a trigonometric polynomial in $\theta$. Even so, equation~\eqref{eq:splitting} is \emph{formal} in the sense that the integrands are defined for $\tau\gg 0$ only up to an arbitrarily high order in $\epsilon$. However, it is \emph{asymptotic}, which is to say that at each order the identity is \emph{exact}. Put differently, \eqref{eq:splitting} is a collection of exact identities, one for each order in $\epsilon$, written in closed form. Moreover, at each order, the integrands are \emph{analytic functions} by virtue of the extension result in Theorem~\ref{thm:extension}. Of course, we need to check that \emph{each order of} \eqref{eq:problem-integral}, \ie, the integral $\regint_{t}^\infty \bigl(\de_\theta f^u_\psi\bigr)^\ell(\tau;z,\theta)\,d\tau$ with arbitrary $\ell\in\N$, exists \footnote{\ldots and this is the second.}.

The point of the formula above is twofold. First, the integral $\regint_{-\infty}^\infty \de_\theta f^u_\psi(\tau)\,d\tau$ (at each order in $\epsilon$) turns out to be exponentially small in the limit $g\to 0$. Second, we can actually construct a (formal) recursion relation for the function $\de_\theta(X^s-X^u)(\tau)$, taking us to ever-increasing orders in $\epsilon$, as follows. Differentiating both sides of \eqref{eq:eqm} with respect to $\theta$ one obtains
\beqn
\begin{pmatrix}
L_t & 0 \\
0 & \de_t^2
\end{pmatrix}
\de_\theta X^{s,u}=
M^{s,u}
\de_\theta X^{s,u}+\lambda f^{s,u}_{\varphi\psi}
\eeqn
with $L_t=\de_t^2-\gamma^2\cos\Phi^0(z e^{\gamma t})$, as in Proposition~\ref{prop:Kinverse}, and
\beqn
M^{s,u}(t;z,\theta)\defas 
\begin{pmatrix}
g^2\cos\Phi^{s,u}(t;z,\theta)-\gamma^2\cos\Phi^0(z e^{\gamma t}) & 0 \\
0 & 0
\end{pmatrix}
+\lambda f^{s,u}_{\varphi \varphi}(t;z,\theta)
.
\eeqn
Owing to the $\theta$-derivative acting on $X^{s,u}$ and Proposition~\ref{prop:Kinverse}, we get
\beqn
\de_\theta X^{s,u}(t;z,\theta)=
\regint_{\pm\infty}^t K(t,t';z)\,\bigl[M^{s,u}
\de_\theta X^{s,u}+\lambda f^{s,u}_{\varphi\psi}\bigr](t';z,\theta)\,d t',
\eeqn 
upon choosing $\infty$ in conjunction with $X^s$ and $-\infty$ with $X^u$. The most convenient way to see this is to infer the identity for $\de_\theta X^u$ and employing $K(-t,-\tau;z\inv)\equiv -K(t,\tau;z)$ and $X^s(t;z,\theta)\equiv (2\pi,0)-X^u(-t;z\inv,-\theta)$. 

At $(z,\theta)=(1,0)$, $M^s(t)\equiv M^u(t)$ and $f^{s}_{\varphi\psi}(t)\equiv f^{u}_{\varphi\psi}(t)$, such that (2) of Proposition~\ref{prop:regint} validates
\beq\label{eq:recursion}
\begin{split}
\de_\theta(X^s-X^u)(\tau) = \,\, &\regint_{\infty}^{-\infty} K(\tau,\tau')\,\bigl[M^u\de_\theta X^u+\lambda f^{u}_{\varphi\psi} \bigr](\tau')\,d\tau'\,+  \\ 
& +\regint_{\infty}^\tau K(\tau,\tau')\,\bigl[M^s\de_\theta(X^s-X^u)\bigr](\tau')\,d\tau'.
\end{split}
\eeq
By construction, the quantities appearing in square brackets above are $\theta$-gradients:
\beqn
M^{s,u}\de_\theta X^{s,u}+\lambda f^{s,u}_{\varphi\psi}=
\de_\theta\mathfrak{A}^{s,u}
\eeqn
with, for instance \footnote{We could make $\mathfrak{A}^{s,u}$ of order $\epsilon$ by deleting its lowest order, because the latter does not depend on $\theta$.}, 
\beqn
\mathfrak{A}^{s,u} \defas
\begin{pmatrix}
L_t & 0 \\
0 & \de_t^2
\end{pmatrix}\!
X^{s,u}
=
\Omega(X^{s,u})-
\begin{pmatrix}
\gamma^2\cos\Phi^0(z e^{\gamma t}) & 0 \\
0 & 0
\end{pmatrix}\!
X^{s,u},
\eeqn
regardless of the value of $(z,\theta)$. In addition, but only at $(z,\theta)=(1,0)$,
\beqn
\bigl[M^s\de_\theta(X^s-X^u)\bigr](t)
=\bigl[\de_\theta(\mathfrak{A}^s-\mathfrak{A}^u)\bigr](t).
\eeqn

Notice that \eqref{eq:recursion} is a fixed point equation of the form $\zeta=\E+BM^s\zeta$ solved formally by $\zeta(t)=\de_\theta(X^s-X^u)(t)$. This is highly interesting from the point of view of asymptotic analysis, since $M^s=\order{\epsilon}$ multiplies $\zeta$ on the right-hand side. Indeed, truncating the Taylor series in $\epsilon$, \eg, 
$
\zeta^{\leq k}=\sum_{i=0}^k\epsilon^i\zeta^i, 
$
the identities $\zeta^{k}=(\E+BM^s\zeta)^{k}=(\E+BM^s\zeta^{\leq k-1})^{k}$ become exact. Hence, we can recursively construct any order of $\de_\theta(X^s-X^u)(t)$ from its lower orders, without the need of diverting from $\{(t;z,\theta)=(t;1,0)\,|\,t\in\R\}$ in order to compute the $\theta$-derivative. Cumulatively, we have $\zeta^{k}=\bigl[\sum_{j=0}^{k-1}(BM^s)^j \E\bigr]^{k}$, where $j$ is a power.

\begin{proposition}\label{prop:Neumann-asympt}
In brief, the (presumably) divergent Neumann series $\sum_{j=0}^{\infty}(BM^s)^j \E$ is an asymptotic expansion of $\de_\theta(X^s-X^u)(t)$ in the sense that
\begin{enumerate}
\item At each order in $\epsilon$ it terminates after finitely many well-defined terms, and
\item $\Bigl[\de_\theta(X^s-X^u)-\bigl[\sum_{j=0}^{\infty}(BM^s)^j \E\bigr]^{\leq k}\Bigr](t)=\order{\epsilon^{k+1}}$ for each $k\in\N$.
\end{enumerate}
\end{proposition}
Once inserted into \eqref{eq:splitting}, the latter series provides us with an \emph{asymptotic expansion for the splitting matrix $\Upsilon$ in \eqref{eq:splitting-homoclinic}}:
\begin{corollary}\label{cor:almost-there}
In the asymptotic sense of Proposition~\ref{prop:Neumann-asympt},
\beqn
\Upsilon(t)=\lambda\,\regint_{-\infty}^\infty \de_\theta f^u_\psi(\tau)\,d\tau +\lambda\,\regint_{t}^\infty \Bigl[f^s_{\psi\varphi}\sum_{j=0}^{\infty}(BM^s)^j \E\Bigr](\tau)\,d\tau.
\eeqn
\end{corollary}

The operator $B$ above has the expression
\beq\label{eq:Bexpression}
Bh(t;z,\theta)=\regint_{\infty}^t K(t,\tau;z)h(\tau;z,\theta)\,d\tau,
\eeq
whereas $\E$ is the restriction of
$
\E(t;z,\theta)\defas\regint_{\infty}^{-\infty}K(t,\tau;z)\,\de_\theta\mathfrak{A}^u(\tau;z,\theta)\,d\tau
$
to $(z,\theta)= (1,0)$. 

Using \eqref{eq:Ksplit}, also $\E$ splits into pieces:
$
\E(t;z,\theta)=\sum_{i=0}^1 \sum_{p=0}^i\sum_{j=1}^2 t^p K_{ij}(t;z)\,\E_{ij}^p(z,\theta),
$
where 
\beqn
\E_{ij}^p(z,\theta)\defas -\regint_{-\infty}^{\infty}(-\tau)^{\delta_{i1}-p}\bar K_{ij}(\tau;z)\,\de_\theta\mathfrak{A}^u(\tau;z,\theta)\,d\tau.
\eeqn
We shall see that these \emph{$t$-independent factors are exponentially small} with respect to $g$, as the latter tends to zero. Furthermore, denoting
$
K_{ij}^p(t;z)\defas t^p K_{ij}(t;z),
$
\beqn
(BM^s)^l\E=\sum_{i=0}^1 \sum_{p=0}^i\sum_{j=1}^2 \bigl[(BM^s)^l K_{ij}^p\bigr]\,\E_{ij}^p.
\eeqn 
This is so because, by Proposition~\ref{prop:regint}, all integrals due to \eqref{eq:Bexpression} \emph{factorize} formally: if $h$ is a trigonometric polynomial at each order, then 
\beqn
\regint_t^\infty (h\E)(\tau;z,\theta)\,d\tau=\sum_{i=0}^1 \sum_{p=0}^i\sum_{j=1}^2 \biggl[\,\regint_t^\infty h(\tau;z,\theta)\tau^p K_{ij}(\tau;z)\,d\tau\biggr] \E_{ij}^p(z,\theta).
\eeqn

By virtue of Corollary~\ref{cor:almost-there}, we infer
\begin{proposition}\label{prop:asymptotic}
On the homoclinic trajectory, \ie, setting $(z,\theta)=(1,0)$, the following asymptotic expansion of the splitting matrix holds:
\beq\label{eq:as-expansion}
\Upsilon(t)=\regint_{-\infty}^\infty \lambda\,\de_\theta f^u_\psi(\tau)\,d\tau +\lambda \, c_{ij}^p(t) \E_{ij}^p, 
\eeq
where repeated indices are contracted ($i\in\{0,1\}$, $p\in\{0,i\}$, $j\in\{1,2\}$) and
\beqn
c_{ij}^p(t)\defas \regint_{t}^\infty \Bigl[f^s_{\psi\varphi}\sum_{l=0}^{\infty}(BM^s)^l K_{ij}^p\Bigr](\tau)\,d\tau.
\eeqn
\end{proposition}
As already pointed out, the terms appearing in the asymptotic expansion of Proposition~\ref{prop:asymptotic} will turn out exponentially small with respect to $g$. The factors $c_{ij}^p$ shall pose no problems, the functions $M^s$ and $K_{ij}^p$ being explicit and simple. In brief, Theorem~\ref{thm:splitting} begins to emerge!

For the record, writing $B^uh(t;z,\theta)\defas \regint_{-\infty}^t K(t,\tau;z)h(\tau;z,\theta)\,d\tau$, we get (at $(z,\theta)=(1,0)$)
\beq\label{eq:c_ij}
c_{ij}^p(t)=(-1)^{p+j\delta_{i0}}\,\regint_{-\infty}^{-t}\Bigl[f^u_{\psi\varphi}\sum_{l=0}^{\infty}(B^uM^u)^l K_{ij}^p\Bigr](\tau)\,d\tau.
\eeq

\subsection{Emergence of exponential smallness}\label{subsec:exp-small}

The ``asymptotic'' integrals in \eqref{eq:as-expansion}, including $\lambda\,\E_{ij}^p$, are of the form
\beqn
\regint_{-\infty}^\infty \de_\theta F(X^u;t;z,\theta)\,dt \equiv \sum_{\ell=1}^\infty\epsilon^\ell\sum_{0<\abs{q}\leq \ell N} iq\, e^{iq\cdot \theta}\,\regint_{-\infty}^\infty \bigl[\hat F(\hat X^u;t;z,q)\bigr]^\ell\,dt,
\eeqn
evaluated at $(z,\theta)=(1,0)$, where the integrand on the right-hand side is a trigonometric polynomial of degree $\leq\ell N$ by Theorem~\ref{thm:extension}. We also point out that, by construction, the latter are $\theta$-gradients, which allows us to omit the harmful $q=0$ terms. Above, F depends on $X=X^u$ locally, \ie, only through $X(t;z,\theta)$, as well as analytically near $X^0$, \ie, the series
\beq\label{eq:Fexpansion}
F(X)=F(X^0+\widetilde X)=\sum_{k=0}^\infty F^{(k)} \bigl(\widetilde X\bigr)^{\otimes k}
\eeq
converges. In fact, since $F$ is one of the functions in
\beq\label{eq:Fcollection}
\bigl\{\lambda f_\psi^u\bigr\}\, \bigcup\, \bigl\{-\lambda(-t)^{\delta_{i1}-p}\bar K_{ij}\,\mathfrak{A}^u \;\big|\;\text{$0\leq p\leq i\leq 1$ and $1\leq j\leq 2$}\bigr\},
\eeq
we observe that
\beq\label{eq:F-identity}
F(X^u;t;z,\theta)\equiv t^p F(X^u;0;ze^{\gamma t},\theta+\omega t)\qquad(p\in\{0,1\}),
\eeq
with $(z,\theta)\mapsto [F(X^u;0;z,\theta)]^\ell$ analytic on $(\mathbb{U}_{\tau,\vartheta}\nonzero)\times\{\abs{\impart\theta}\leq\sigma\}$ for all $\ell$ by virtue of Theorem~\ref{thm:extension}. At $z=0$ there is a simple pole, due to $\bar K_{02}$, at worst.

We use the following lemma for analyzing such integrals. Its proof is given in Appendix~\ref{app:computations}. 
\begin{lemma}[Shift of contour]\label{lem:shift-of-contour}
Suppose that the function $h(t;z,\theta) \linebreak[0] \equiv t^p h(ze^{gt},\theta+\omega t)$ is analytic with respect to $(z,\theta)\in(\mathbb{U}_{\tau,\vartheta}\nonzero)\times\{\abs{\impart\theta}\leq\sigma\}$ and $\hat h(\piste,0)=0$. Moreover, set $t_q\defas \sgn(\omega\cdot q)\vartheta g\inv$ and
\beq\label{eq:H_q}
H_q(R)\defas\int_0^\infty e^{(iq\cdot\omega-R)t}(t+it_q)^p\hat h(e^{g(t+it_q)},q)\,dt
\eeq
for each $q\in\Z^d\nonzero$. If
\beqn
\biggl |\, \regint_{-\infty}^0 e^{iq\cdot\omega t}(t+it_q)^p\hat h(e^{g(t+it_q)},q)\,dt\biggr | \leq  A(g)e^{-\sigma\abs{q}}
\eeqn
and if $H_q(R)$ admits an analytic continuation to $\{0<\abs{R}\leq \rho\}$ with a pole of order $k$ at $R=0$, respecting the bound
\beqn
\sup_{\abs{R}=\rho}\abs{H_q(R)}\leq B(g)e^{-\sigma\abs{q}},
\eeqn
then we obtain the exponentially small ($c>0$) bound 
\beqn
\biggl| \,\regint_{-\infty}^\infty h(t;1,0)\,dt\biggr|\leq C\biggl[A(g)+B(g)\sum_{j=0}^k \frac{1}{j!}\Bigl(\frac{\rho\vartheta}{g}\Bigr)^j\biggr]e^{-cg^{-1/(\nu+1)}}.
\eeqn
\end{lemma}

With the aid of Lemma~\ref{lem:shift-of-contour}, we shall prove in Section~\ref{sec:splitting-proof} the following key result:
\begin{proposition}[Convergence vs. exponential smallness]\label{prop:dichotomy} Fix a $t\in\R$. There exist positive constants $c$, $\epsilon_1$ and $C$, such that the estimates
\beqn
|\Upsilon^\ell(t)|\leq 
C\begin{cases}
\epsilon_1^{-\ell}\ell!^{4(\nu+1)} e^{-cg^{-1/(\nu+1)}}\\
\epsilon_1^{-\ell}
\end{cases}
\eeqn
both hold true for all $\ell=1,2,\ldots$.
\end{proposition}
This dichotomy, in which the exponential smallness competes with the usual bound due to convergence, is not new; see \cite{GallavottiTwistless,GallavottiSplit,Procesi}. Remarkably, we get the same exponent of the factorial, $4(\nu+1)$, as the latter articles---even though our method is quite different. 

Theorem~\ref{thm:splitting} follows immediately from Proposition~\ref{prop:dichotomy} by an argument due to Gallavotti, \textit{et al.}: For each $g$, let $n(g)$ be a positive integer. If $\abs{\tilde\epsilon}<\half$ and $\epsilon=\tilde\epsilon\epsilon_1 n(g)^{-4(\nu+1)}$,
\begin{align*}
|\Upsilon(t)| & \leq C e^{-cg^{-1/(\nu+1)}}\sum_{\ell=1}^{n(g)}\Bigl(\frac{\abs{\epsilon}}{\epsilon_1}\Bigr)^\ell\ell!^{4(\nu+1)}+ C\sum_{\ell=n(g)+1}^\infty \Bigl(\frac{\abs{\epsilon}}{\epsilon_1}\Bigr)^\ell \\
& \leq C e^{-cg^{-1/(\nu+1)}}\frac{\abs{\epsilon}}{\epsilon_1} n(g)^{4(\nu+1)}+C\Bigl(\frac{\abs{\epsilon}}{\epsilon_1}\Bigr)^{n(g)+1} \\
& \leq C \abs{\tilde\epsilon}e^{-cg^{-1/(\nu+1)}}+C \abs{\tilde \epsilon} e^{n(g)\ln\abs{\tilde\epsilon}},
\end{align*}
since $\ell!\leq \ell^\ell\leq n(g)^\ell$ in the first sum. Now, set $n(g)=cg^{-1/(\nu+1)}/\ln 2$, such that $n(g)\ln\abs{\tilde\epsilon}\leq -cg^{-1/(\nu+1)}$ and $
\tilde\epsilon=(\epsilon/\epsilon_1) (c/\ln 2)^{4(\nu+1)}g^{-4}$ (demanding in particular that $\epsilon g^{-4}$ be small).
\begin{remark}
We used the exponentially small but diverging estimate to bound the partial sum $\sum_{\ell=1}^{n(g)}\epsilon^\ell\Upsilon^\ell$, whereas the remainder of the series was easily controlled by convergence. As $n(g)\to\infty$ with $g\to 0$, the important thing here is to have the exponentially small bound on $\Upsilon^\ell$ for \emph{arbitrarily} large $\ell$, in addition to the $\epsilon$-analyticity of $\Upsilon$.
\end{remark}

\section{Proof of Theorem~\ref{thm:splitting}}\label{sec:splitting-proof}

We are left with proving Proposition~\ref{prop:dichotomy}, since Theorem~\ref{thm:splitting} was already shown to be its corollary. Here things are most conveniently explained using tree diagrams. However, each tree will be treated as an \emph{individual}, solely for bookkeeping benefits, and no cancellations nor regroupings are forced upon them.

By Proposition~\ref{prop:asymptotic}, we need to consider the simple factors $c_{ij}^p$, as well as the integrals
\beq\label{eq:Fintegral}
\regint_{-\infty}^\infty \hat F(\hat X^u;t,z,q)\,dt\equiv \sum_{\ell=1}^\infty\epsilon^\ell\,\regint_{-\infty}^\infty \bigl[\hat F(\hat X^u;t;z,q)\bigr]^\ell\,dt
\eeq
of Subsection~\ref{subsec:exp-small}, at each order $\ell\geq\abs{q}/N>0$. As $F$ comes from the collection in \eqref{eq:Fcollection}, all integrals of the latter type shall be controlled with the aid of Lemma~\ref{lem:shift-of-contour}.

Due to the superscript $u$---referring to the unstable manifold---in the integrand above, the required bounds on the integrals over $\R_-$ in $\regint_{-\infty}^\infty=\regint_{-\infty}^0+\regint_{0}^\infty$ are straightforward, and are discussed later.

In order to deal with $\regint_0^\infty$, we present the procedure below, which amounts to \emph{little more than integration by parts}. First, we expand $F$ according to \eqref{eq:Fexpansion} and, like in Section~\ref{sec:continuation}, split
\beqn
\widetilde X=\widetilde X_{\leq 1}+\delta_2 \widetilde X=\,\raisebox{3pt}{\pstree{\Tp}{\TC*}} \, + \, \raisebox{3pt}{\pstree{\Tp}{\Tc{\rad}}} \; ,
\eeqn 
dropping the superscript $u$ from the notation. We can then express $F$ pictorially as
\vspace{-2mm}
\beqn
\sum_{m=0}^\infty\,\sum_{0\leq m'\leq m}\binom{m}{m'}\raisebox{3pt}{
\pstree{\Tp}{
  \pstree[treesep=3mm]{\TC*[name=n0]}{
    \Tc[name=n1]{\rad} \Tc[name=n2]{\rad} \TC*[name=n3] \TC*[name=n4]}}
\ncarc[linestyle=dotted,nodesep=1pt]{n1}{n2}\trput{\tiny $m'$}
\ncarc[linestyle=dotted,nodesep=1pt]{n3}{n4}\trput{\tiny $m-m'$}
\nput[labelsep=1mm]{115}{n0}{\tiny $F^{(m)}$}
}\qquad,
\eeqn
where the binomial coefficient comes from the combinatorics of shuffling the arguments of the symmetric $F^{(m)}$ \footnote{This convention merely facilitates drawing.}, which is attached to the root (node). Next, we replace the subtrees $\raisebox{3pt}{\pstree{\Tp}{\Tc{\rad}}}=\delta_2 \widetilde X$ with the expansion \eqref{eq:tree-exp2} derived from \eqref{eq:tree-rec}.

In a tree $T_{v_0}$ with root $v_0$, a generic subtree $T_v$ with root $v$ has then the expression
\beq\label{eq:subtree}
T_v=U_v\bigl(\K\inv T_{w_1},\dots,\K\inv T_{w_{m_v'}};\bigl(\widetilde X_{\leq 1}\bigr)^{\otimes (m_v-m'_v)}\bigr),
\eeq
where $T_{w_j}$ is a subtree---with root $w_j$---entering $v$ and the \emph{node function} 
\beqn
U_v=
\begin{cases}
\text{$F^{(m_v)}$ coming from \eqref{eq:Fcollection}}, & \text{if $v=v_0$}, \\
\text{$w^{(m_v)}$ as in \eqref{eq:wdefinition}}, & \text{otherwise}. 
\end{cases}
\eeqn
In a pictorial representation, there are $m_v$ lines entering the node $v$ and precisely $m_v-m_v'$ of the latter are leaving a black dot (\raisebox{3pt}{\TC*}) end node:
\vspace{-2mm}
\beqn
\raisebox{3pt}{
\pstree{\Tp}{\tiny \Tcircle{\text{$T_{v}$}}}
}
=
\raisebox{3pt}{
\pstree{\Tp}{
  \pstree[treesep=3mm]{\TC*[name=n0]}{
    \Tcircle[name=n1]{\tiny \text{$T_{w_1}$}} \Tcircle[name=n2]{\tiny \text{$T_{w_{m_v'}}$}} \TC*[name=n3] \TC*[name=n4]}}
\ncarc[linestyle=dotted,nodesep=1pt]{n1}{n2}\naput[nrot=0]{\tiny $m'_v$}
\ncarc[linestyle=dotted,nodesep=1pt]{n3}{n4}\trput{\tiny $m_v-m'_v$}
\nput[labelsep=1mm]{115}{n0}{$U_v$}
}\qquad.
\eeqn 

The ``whole'' tree $T_{v_0}$ contributes at orders $\ell$ satisfying
\beq\label{eq:ell-bound}
\ell\geq 1+(m_{v_0}-m'_{v_0})+\sum_{j=1}^{m_{v_0}'}\deg{T_{w_j}}\defasr \operatorname{d}(T_{v_0}),
\eeq
where $\deg{T_w}$---defined in \eqref{eq:deg}---counts end nodes with suitable weights as well as nodes with exactly one entering line in the subtree $T_w$, and the constant term $1$ counts the root ($F^{(m_{v_0})}=\order{\epsilon}$). By this we mean that $\operatorname{d}(T_{v_0})$ is the largest integer such that
\beqn
T_{v_0}=\order{\epsilon^{\operatorname{d}(T_{v_0})}}\quad\text{as}\quad \epsilon\to 0.
\eeqn

\subsection{Some combinatorics}
\begin{lemma}\label{lem:number-of-nodes}
The number of nodes, $\operatorname{n}(T_{v_0})$, of a tree $T_{v_0}$ contributing at order $\ell\geq 2$ obeys
\beq\label{eq:number-of-nodes}
\operatorname{n}(T_{v_0})\leq 2\bigl(\operatorname{d}(T_{v_0})-1\bigr)\leq 2(\ell-1).
\eeq
The number of end nodes is at most $\operatorname{d}(T_{v_0})-1\leq \ell-1$.
\end{lemma}
\begin{proof}
$\operatorname{n}(T_{v_0})$ attains its maximum with respect to the ``degree'' $\operatorname{d}(T_{v_0})$ as follows: 
\begin{enumerate}
\item If $\operatorname{n}(T_{v_0})$ is even, there is only one line entering the root $v_0$ and the rest of the tree is binary, \ie, contains only end nodes and nodes with exactly two entering lines. 
\item If $\operatorname{n}(T_{v_0})$ is odd, the tree is binary.
\end{enumerate}
Moreover, each of the end nodes is either \raisebox{3pt}{\TC*} or \raisebox{3pt}{\Tcircle{\tiny $1$}}, which contribute the least to the degree; see \eqref{eq:deg}. These choices minimize the number of end nodes when the $\order{\epsilon}$ nodes having exactly one entering line are excluded (except at $v_0$ which is always $\order{\epsilon}$). Therefore, $\operatorname{d}(T_{v_0})$ gets minimized with respect to the number of all nodes, $\operatorname{n}(T_{v_0})$. Since in a binary tree of $j$ end nodes there are $2j-1$ nodes, we infer \eqref{eq:number-of-nodes}.

The root and each of the end nodes increases $\operatorname{d}(T_{v_0})$ by at least one, which implies the bound on the number of end nodes. 
\end{proof}

\begin{corollary}\label{cor:number-of-trees}
At most $2^{6\ell}$ trees contribute at order $\ell$.
\end{corollary}
\begin{proof}
It is well-known that the number of (rooted) trees with $k$ indistinguishable nodes is
$
N(k)\defas\frac{1}{k}\binom{2k-2}{k-1}\leq \frac{1}{k}4^{k-1},
$
which follows from generating functions (see \cite{Drmota}) and the bound $(2m)!\leq 4^m(m!)^2$. By Lemma~\ref{lem:number-of-nodes}, the number of end nodes is less than $\ell$. We label the latter arbitrarily by the labels in $\{\raisebox{3pt}{\TC*}\}\bigcup \bigl\{\raisebox{3pt}{\Tcircle{\text{\tiny $m$}}}\;|\; \text{$1\leq m\leq\ell-1$}\bigr\}$ in order to form an upper bound on the number of our trees; these are the only possible labels, as otherwise $\operatorname{d}(T_{v_0})$ certainly exceeds the order $\ell$---contradicting \eqref{eq:ell-bound}. The labeling of a tree with $j$ end nodes can be carried out in at most $\binom{j+\ell-1}{j-1}\leq 2^{j+\ell-1}$ ways. The desired bound is thus obtained from
$
1+\sum_{k=2}^{\ell-1}N(k)\binom{\ell+k-2}{k-2}+\binom{2\ell-2}{\ell-2}\sum_{k=\ell}^{2(\ell-1)}N(k)\leq 2^{6\ell},
$
because by Lemma~\ref{lem:number-of-nodes} the number of end nodes is at most $\ell-1$ even though the total number of nodes can be as large as $2(\ell-1)$. The term 1 on the left-hand side counts the single node tree $F^{(0)}$.
\end{proof}

\subsection{Simplification of integrals: scalar trees}

We take a preliminary step towards bounding the values of the trees.

Let us split the kernels $K$ of the operators $\K\inv$ appearing in \eqref{eq:subtree} into four pieces according to \eqref{eq:Ksplit}. These operators are attached to the lines between the nodes of a tree. Each of the $2^{6\ell}$ trees counted in Corollary~\ref{cor:number-of-trees} thus breaks into at most $4^{2\ell}$ new trees, as there are no more than $2\ell$ such lines by Lemma~\ref{lem:number-of-nodes}. 

At the same time, we also expand the matrix products due to the coordinate representation of \eqref{eq:subtree} at each node:
$
(U_v)^i(\widetilde T_1,\dots,\widetilde T_m)=\sum_{j_1=1}^{d+1}\dots\sum_{j_m=1}^{d+1} (U_v)^i_{j_1\dots j_m}\widetilde T_1^{j_1}\dots \widetilde T_m^{j_m},
$
where the $\widetilde T_k$'s represent all arguments of $U_v$ (\ie, subtrees entering the node $v$)---including $\widetilde X_{\leq 1}$---and the superindices specify vector components. We next separate each scalar term into its own tree, thus getting up to $(d+1)^{2\ell}$ of these \emph{scalar trees} from each old tree. 

There are lines carrying a factor $t-\tau$, coming from the $i=1$ terms of \eqref{eq:Ksplit}. They correspond to double integrals: $\iint_{-\infty}^t (d\tau)^2 =\int_{-\infty}^t d\tau\,(t-\tau)$. We remove these factors, insert a new node $\tilde v$ on the line with a node function $U_{\tilde v}\equiv 1$ and an integral sign both on the line leaving and on the line entering $\tilde v$. This operation can be depicted as
\vspace{1mm}
\beq\label{eq:iint}
\raisebox{3pt}{
\pstree[levelsep=1.3cm]{\Tp}{\Tp \tbput{$\iint_{-\infty}^t$}}
} \quad \longmapsto \quad
\raisebox{3pt}{
\pstree[levelsep=1.3cm]{\Tp}{
  \pstree{\TC*[name=v] \tbput{$\int_{-\infty}^t$}}{\Tp \tbput{$\int_{-\infty}^\tau$}}}
\nput[labelsep=1mm]{100}{v}{$1$}
}.
\vspace{4mm}
\eeq
It does not affect the number of trees, but slightly simplifies the discussion below, even though the number of lines in a single tree can be as much as doubled.

Altogether, we arrive at the following conclusion:
\begin{proposition}\label{prop:tree-count}
There are less than $C_d^{\ell}$ scalar trees to be considered, with $C_d=2^{10}(d+1)^2$, at each order $\ell$. Each of these trees has at most $4(\ell-1)$ lines.
\end{proposition}

\begin{remark}[Some conventions]\label{rem:convention}
From now on, by a tree we will always refer to a scalar tree, where all the decompositions above have been carried out. In order to alleviate notation, we systematically omit all vector component indices.
\end{remark}

Recall that the node $w'$ is the unique predecessor of the node $w$. Let $T_{v_0}$ be a generic (scalar) tree with root $v_0$. Denote by $V$ the set of all nodes and by $V_\text{int}$ the set of \emph{integrated nodes}, \ie, the nodes whose leaving line carries an integral. We consider the root $v_0$ an integrated node, such that $V_{\text{int}}$ consists of all nodes of $T_{v_0}$ except black dot (\raisebox{3pt}{\TC*}) end nodes. We can describe $T_{v_0}$ by giving its structure recursively: if $T_v$ is the subtree of $T$ with root $v\in V_{\text{int}}$, then
\beq\label{eq:scalar-subtree}
T_v(t;z,\theta)=u_v(t;z,\theta)\prod_{\substack{w\in V_{\text{int}}\\w'=v}} \int_{-\infty}^{t} T_w(\tau;z,\theta)\,d\tau.
\eeq
We call $u_v$ the \emph{multiplier} of the node $v$. It is a scalar function comprising all factors in the expression of the tree carrying the same variable of integration (``time label''), constricted in-between integral signs. In particular, it contains all subtrees $\raisebox{3pt}{\pstree{\Tp}{\TC*}}$ entering $v$, as these are the functions $\widetilde X_{\leq 1}$ involving no integrals.

To be completely explicit, the multiplier $u_v$ of a node $v\in V_{\text{int}}$ is one of the functions below:
\begin{enumerate}
\item\label{list:multipliers} $\bar K_v\delta_2h^{(k_v)}$, if $v$ is the end node \raisebox{3pt}{\Tcircle{\tiny $k_v$}}; see \eqref{eq:endnode}.
\item $\bar K_v w^{(m_v)}\Bigl(\prod_{\substack{w\notin V_{\text{int}}\\w'=v}}\widetilde X_{\leq 1}\Bigr) \Bigl(\prod_{\substack{w\in V_{\text{int}}\\w'=v}}K_{w}\Bigr)$, if $v$ is neither an end node nor the root $v_0$, and didn't appear from splitting a double integral according to \eqref{eq:iint}.
\item\label{item:double-int-split} $1$, if $v$ appeared from splitting a double integral; see \eqref{eq:iint}.
\item\label{item:root-t} $F^{(m_v)}\Bigl(\prod_{\substack{w\notin V_{\text{int}}\\w'=v}}\widetilde X_{\leq 1}\Bigr)\Bigl(\prod_{\substack{w\in V_{\text{int}}\\w'=v}}K_{w}\Bigr)$, if $v=v_0$; see \eqref{eq:Fcollection}.
\end{enumerate}
The functions $K_v$ and $\bar K_v$ refer to diagonal elements of $K_{ij}$ and $\bar K_{ij}$ in \eqref{eq:Ksplit}, respectively. $m_v'$ is the cardinality of $\{w\in V_\text{int}\:|\:w'=v\}$ and $m_v-m_v'$ the cardinality of $\{w\in V\setminus V_\text{int}\:|\:w'=v\}$. 

We draw the scalar (sub)tree in \eqref{eq:scalar-subtree}---originated from \eqref{eq:subtree}---as
\beqn
\raisebox{3pt}{
\pstree{\Tp}{\Tcircle{\tiny \text{$T_{v}$}}}
}
=
\raisebox{3pt}{
\pstree{\Tp}{
  \pstree[treesep=3mm]{\TC*[name=n0]}{
    \Tcircle[name=n1]{\tiny \text{$T_{w_1}$}} \Tcircle[name=n2]{\tiny \text{$T_{w_{m_v'}}$}} }}
\ncarc[linestyle=dotted,nodesep=1pt]{n1}{n2}\naput[nrot=0]{\tiny $m'_v$}
\nput[labelsep=1mm]{115}{n0}{$u_v$}
}\qquad.
\vspace{2mm}
\eeqn 
This diagram is reminiscent of the one below \eqref{eq:subtree}, except that the operator $U_v$ has changed into the multiplier $u_v$ and the subtrees $\raisebox{3pt}{\pstree{\Tp}{\TC*}}\:$ have been absorbed into $u_v$. Moreover, recalling the decompositions above, the present diagram carries $\int_{-\infty}^t$ on its lines instead of $\K\inv$---compare \eqref{eq:scalar-subtree} with \eqref{eq:subtree}---and has possibly more nodes due to the diagram in \eqref{eq:iint}.

We do not consider the end nodes with a black dot (\raisebox{3pt}{\TC*}) nodes anymore. Subsequently, we will refer to ``integrated nodes'' (elements of $V_\text{int}$) as just ``nodes''.

\subsection{Integration by parts: one step}
Reinserting the implicit arguments $(z,\theta)$, 
\beq\label{eq:multiplier-prop}
u_v(t;z,\theta)\equiv t^p u_v(ze^{\gamma t},\theta+\omega t)
\eeq
for the obviously defined $u_v(z,\theta)$. The power $p$ can be nonzero only at the root, $v=v_0$, where it possibly assumes the value 1 due to case (\ref{item:root-t}) in the list of all possible multipliers above.

We now turn our attention to the Fourier transformed integrals 
\beq
\regint_{0}^\infty T_{v_0}(t;z,q)\,dt =\regint_0^{\infty}\,\sum_{q_{v_0}+\sum_{w'=v_0} \tilde q_w=q} u_{v_0}(t;z,q_{v_0})\prod_{w'={v_0}} \int_{-\infty}^{t} T_w(\tau;z,\tilde q_w)\,d\tau\,dt,
\eeq
which---recalling \eqref{eq:scalar-subtree}---arise from \eqref{eq:Fintegral}, at each order $\ell\geq \abs{q}/N>0$. We omit the usual $\widehat{\text{hat}}$ in Fourier transforms, since there is no danger of confusion. Due to Theorem~\ref{thm:extension}, we may impose the finiteness condition
\beq\label{eq:Fourier-condition}
\abs{q_{v_0}} + \sum_{w'=v_0}\abs{\tilde q_w} \leq \ell N,
\eeq 
such that $\regint_{0}^\infty T_{v_0}(t;z,q)\,dt$ agrees at order $\ell$ with
\beq\label{eq:reg-tree-int}
\sideset{}{^\star}\sum_{q_{v_0}+\sum_{w'=v_0} \tilde q_w=q} \res_{R=0}\frac{1}{R}\int_0^{\infty}e^{-Rt}\,u_{v_0}(t;z,q_{v_0})\prod_{w'={v_0}} \int_{-\infty}^{t} T_w(\tau;z,\tilde q_w)\,d\tau\,dt.
\eeq
Here the asterisk reminds us that \eqref{eq:Fourier-condition} is being respected by the sum.

\begin{remark}\label{rem:int-task}
The task is to show that the $t$-integral in \eqref{eq:reg-tree-int} extends analytically from large positive values of $\repart{R}$ to a punctured neighbourhood of $R=0$, such that the residue can be computed. For this, we need the specific structure of the multipliers $u_v$.
\end{remark}

Let us define
$
\xi_{01}=\bar\xi_{02}=1,$
$\bar\xi_{01}=\xi_{02}=\xi_{11}=\bar\xi_{11}=-1,$ and
$\xi_{12}=\bar\xi_{12}=0.
$
There exist positive, continuous, functions $a_{ij}$ and $\bar a_{ij}$ such that
\beq\label{eq:K-bounds}
\abs{K_{ij}(t;z)}\leq a_{ij}(z) e^{\xi_{ij}\gamma\abs{t}}\mathand\abs{\bar K_{ij}(t;z)}\leq \bar a_{ij}(z) e^{\bar\xi_{ij}\gamma\abs{t}}
\eeq
hold on $\{(t,z)\in\R\times \C\;|\;\text{$\abs{z}=1$ and $ze^{\gamma t}\neq\pm i$}\}$.
Moreover,
\beq\label{eq:a-prod}
a_{ij}\bar a_{ij} \leq \frac{A}{\gamma^{1-i}},
\eeq
where the constant $A$ is independent of $\gamma$. Notice that, in the Kronecker delta notation,
\beq\label{eq:xi-sum}
\xi_{ij}+\bar\xi_{ij}=-2\,\delta_{i1}\delta_{j1}\leq 0.
\eeq
Emphasizing the node in question, instead of specifying the sub\-indices $ij$ we write $\xi_v$ and $\bar\xi_v$.  

For each node $v$, we define the numbers
\beqn
r_v\defas\sup_{(z,\theta)\in B}\min{\bigl\{k\in\Z\;\big|\;\forall\,\delta>0:\text{$u_v(t;z,\theta)e^{-(k+\delta)\gamma t}\to 0$ as $t\to\infty$}\bigr\}}
\eeqn
with 
\beq\label{eq:ok-domain}
B\defas \{\abs{z}=1, \abs{\arg{z}}\leq\vartheta\}\times\{\abs{\impart\theta}\leq\sigma\}.
\eeq
Thus, recalling \eqref{eq:multiplier-prop}, we are inside the analyticity domain $\mathbb{U}_{\tau,\vartheta}\times\{\abs{\impart\theta}\leq\sigma\}$ of Lemma~\ref{lem:an-cont}. These numbers measure the divergence rate of the multipliers $u_v$ in the limit $t\to\infty$. 
\begin{lemma}\label{lem:r_v}
Ordered according to the list of possible $u_v$'s on p.~\pageref{list:multipliers},
\beqn
r_v=
\begin{cases}
\bar\xi_v+k_v, & \text{case (1),} \\
\bar\xi_v +(m_v-m_v')+\sum_{w'=v}\xi_w, & \text{case (2),} \\
0, & \text{case (3),} \\
\bar\xi_F+(m_v-m_v')+\sum_{w'=v}\xi_w, & \text{case (4)},
\end{cases}
\eeqn
where $\bar\xi_F\in\{0,1\}$ depends on the choice of $F$ in \eqref{eq:Fcollection}. Moreover $u_v(z,\theta)$, see \eqref{eq:multiplier-prop}, is analytic on $\{\abs{\impart\theta}\leq\sigma\}$ with respect to $\theta$ and on $\mathbb{U}_{\tau,\vartheta}\nonzero$ with respect to $z$. It is also analytic in the punctured neighbourhood $\{\abs{z}\geq \tau\inv\}$ of $z=\infty$, at which point there is a (possible) pole of order $r_v$ (if $r_v>0$). 
\end{lemma}
\begin{proof}
The maps $\Phi^0(z)$ and $f(\Phi^0(z),\theta)$ are analytic in these domains, without singularities at $z=0,\infty$; see \eqref{eq:arctan-sym}. The rest follows by staring at the expression of $u_v$ in each case.
\end{proof}

Starting from the end nodes of a tree---setting $s_v\defas 0$ for them---we recursively define
\beq\label{eq:n_v-def}
s_v\defas \sum_{w'=v}\max(0,n_w) \mathand n_v\defas r_v+s_v.
\eeq
These numbers measure the divergence rate of (sub)trees in the limit $t\to\infty$:
\begin{lemma}
For $T_v$ as in \eqref{eq:scalar-subtree}, and any $\delta>0$, the estimates
\beq\label{eq:n_v-bound}
T_v(t;z,\theta)e^{-(n_v+\delta)\gamma t}\to 0\quad\text{as}\quad t\to\infty,
\eeq
\beq\label{eq:s_v-bound}
\biggl(\,\prod_{w'=v} \int_{-\infty}^{t} T_w(\tau;z,\theta)\,d\tau\biggr) e^{-(s_v+\delta)\gamma t}\to 0\quad\text{as}\quad t\to\infty
\eeq
hold true in the region $B$ of \eqref{eq:ok-domain}.
\end{lemma}
\begin{proof}
Assume that \eqref{eq:n_v-bound} is true for each successor $w$ of $v$ ($w'=v$). By \eqref{eq:scalar-subtree}, we only need to observe that, given $\delta>0$,
$
\bigl |\int_{-\infty}^t T_w(\tau;z,\theta)\,d\tau\bigr |\leq C_1+C_2\int_0^t e^{(n_w+\delta)\gamma \tau}\,d\tau=\order{e^{[\max(0,n_w)+\delta]\gamma t}}
$
in the limit $t\to\infty$. For then \eqref{eq:n_v-def} implies \eqref{eq:s_v-bound} for the node $v$. Now, \eqref{eq:n_v-bound} follows from the definition of $r_v$. Since \eqref{eq:n_v-bound} holds for end nodes ($n_v=r_v$),  induction proves the claim.
\end{proof}

Recalling Remark~\ref{rem:int-task}, let us now come back to the integral in \eqref{eq:reg-tree-int}, setting $v=v_0$, \ie, 
\beq\label{eq:integraali}
\int_0^{\infty}e^{-Rt}\,u_v(t;z,q_v)\prod_{w'=v} \int_{-\infty}^{t} T_w(\tau;z,\tilde q_w)\,d\tau\,dt.
\eeq
An obvious problem is the exponential divergence of the integrand; see \eqref{eq:n_v-bound} and \eqref{eq:s_v-bound}. Our cure is the following. Since $u_v$ is meromorphic at $z=\infty$, by Lemma~\ref{lem:r_v}, we may expand
\beq\label{eq:multiplier-split}
u_v(z,\theta)=\sum_{k=-s}^{r_v} z^k u_{v,k}(\theta)+u_{v,< -s}(z,\theta)
\eeq
for any integer $s\geq -r_v$, with
$
u_{v,< -s}(z,\theta)=\order{z^{-s-1}}$ as $z\to\infty$. Extending \eqref{eq:multiplier-prop}, we write
\beqn
u_{v,k}(t;\theta)\equiv t^p \, e^{k\gamma t}\, u_{v,k}(\theta+\omega t)
\mathand
u_{v,< -s}(t;z,\theta)\equiv t^p \, u_{v,< -s}(ze^{\gamma t},\theta+\omega t),
\eeqn
where $p\in\{0,1\}$ depends on the choice of $F$ in case (\ref{item:root-t}) on p.~\pageref{item:root-t}. In particular, the integral
$
\int_0^{\infty}e^{-Rt}\,u_{v,< -s_v}(t;z,q_v)\prod_{w'=v} \int_{-\infty}^{t} T_w(\tau;z,\tilde q_w)\,d\tau\,dt
$
is convergent for $\repart{R}>-\gamma$, by virtue of \eqref{eq:s_v-bound}, and can be estimated on a circle $\abs{R}=\rho<\gamma$ for the purpose of Lemma~\ref{lem:shift-of-contour}.

The rest of the integral \eqref{eq:integraali} is \emph{integrated by parts}: for $-s_v\leq k\leq r_v$ and sufficiently large positive values of $\repart{R}$,
\beq
\begin{split}
& \int_0^{\infty}  e^{-Rt}\, z^k u_{v,k}(t;q_v) \prod_{w'=v} \int_{-\infty}^{t} T_w(\tau;z,\tilde q_w)\,d\tau\,dt   \\
& = \frac{z^k  u_{v,k}(q_v)}{R_v}\Biggl\{\frac{E_v(0)}{R_v^p} +\sum_{p'=0}^p\frac{1}{R_v^{p-p'}} \int_0^\infty e^{-R_v t}\,t^{p'} \frac{dE_v}{dt}\,dt\Biggr\}, \label{eq:int-by-parts}
\end{split}
\eeq
where 
$
R_v\defas R-k\gamma-i\omega\cdot q_v
$
and
$
E_v(t)\defas E_v(t;z,\{\tilde q_{w} \, |\, w'=v\})\defas \prod_{w'=v} \int_{-\infty}^{t} T_w(\tau;z,\tilde q_w)\,d\tau
$.
But $\frac{dE_v}{dt}=\sum_{w'=v} T_{w}(t)\prod_{{\bar w'=v, \bar w\neq w}}\int_{-\infty}^{t} T_{\bar w}(\tau)\,d\tau $, so that the recursion relation \eqref{eq:scalar-subtree} yields
\begin{align*}
\frac{dE_v}{dt} 
 &=\sum_{w'=v} \sum_{q_{w}+\sum_{\bar w'=w} \tilde q_{\bar w} = \tilde q_{w}} u_w(t;z,q_w) \biggl(\,\prod_{\bar w'=w}\int_{-\infty}^{t} T_{\bar w}(\tau;z,\tilde q_{\bar w})\,d\tau\biggr) \biggl(\, \prod_{\substack{\bar w'=v\\ \bar w\neq w}}\int_{-\infty}^{t} T_{\bar w}(\tau;z,\tilde q_{\bar w})\,d\tau \biggr).
\end{align*} 
We now fix $w$, and collect all the Fourier sums together so that they run over the set specified by
$
q_{v}+q_{w} + \sum_{\bar w'=w}\tilde q_{\bar w} + \sum_{\bar w'=v, \bar w\neq w}\tilde q_{\bar w} = q.
$
Again, without affecting the $\ell$th order,  we also impose the finiteness condition 
$
|q_{v}|+|q_{w}| + \sum_{\bar w'=w} | \tilde q_{\bar w}| + \sum_{\bar w'=v, \bar w\neq w} | \tilde q_{\bar w}| \leq \ell N
$
similar to \eqref{eq:Fourier-condition}, which allows us to bring the sum out of the $t$-integral.
We observe that the remaining integral in \eqref{eq:int-by-parts} produces integrals similar to the original \eqref{eq:integraali}, with the following changes:
(i) $u_v$ changes to $u_w$,
(ii) $R$ changes to $R_v$,
(iii) $p$ changes to $p'$, and
(iv) the integral on the line \emph{leaving} $w$, $\int_{-\infty}^{t} T_{w}(\tau)\,d\tau$, is replaced by the ones on the lines \emph{entering} $w$, the product $\prod_{\bar w'=w}\int_{-\infty}^{t} T_{\bar w}(\tau)\,d\tau$.

Thus, a single step in the integration-by-parts scheme can be described in terms of trees as follows: Given a tree, consider one of the successors, $w$, of the root, $v=w'$. The line from $w$ to $v$ is ``contracted'' by erasing the node $w$, reattaching to $v$ all subtrees originally entering $w$, and replacing the multiplier of the root, $v$, by $u_w$. Finally, we rename the root $w$. Pictorially,
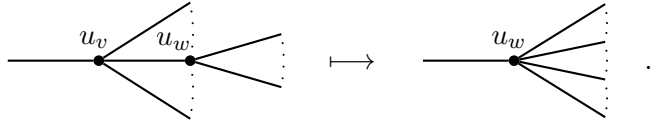
\beqn
\raisebox{3pt}{
\pstree[treesep=0.7cm]
{\Tp}{\pstree{\TC*[name=v]}{\Tp[name=top] \pstree{\TC*[name=cen,bbh=2pt]
}
{\Tp[name=centop] \Tp[name=cenbot]} \Tp[name=bot]}}
\ncarc[linestyle=dotted,nodesep=3pt]{top}{cen}
\ncarc[linestyle=dotted,nodesep=3pt]{cen}{bot}
\ncarc[linestyle=dotted,nodesep=3pt]{centop}{cenbot} 
\nput[labelsep=1mm]{100}{v}{$u_{v}$}
\nput[labelsep=1mm]{120}{cen}{$u_{w}$}
}
\quad \longmapsto \quad
\raisebox{3pt}{
\pstree[treesep=0.5cm]
{\Tp}{\pstree{\TC*[name=v]}{\Tp[name=top1] \Tp[name=centop1] \Tp[name=cenbot1] \Tp[name=bot1]}}
\ncarc[linestyle=dotted,nodesep=3pt]{top1}{centop1}
\ncarc[linestyle=dotted,nodesep=3pt]{centop1}{cenbot1}
\ncarc[linestyle=dotted,nodesep=3pt]{cenbot1}{bot1} 
\nput[labelsep=1mm]{100}{v}{$u_w$}
}\quad .
\eeqn

\subsection{Integration by parts: exhausting the entire tree} 
We can proceed recursively and repeat the procedure in the previous subsection until there are no nodes left in the tree. Start with a tree $T_v$ having root $v$. First, call $T_0\defas T_v$ and set $v_0\defas v$. Then choose a successor $w$ of $v$ and define $v_1\defas w$, contracting the line from $w$ to $v$. Next, in the \emph{new tree} (the rightmost diagram above) called $T_1$, choose a successor of $w$ and call it $v_2$. Contract the line from $v_2$ to $v_1$. Repeat until the tree has been exhausted and all nodes have been numbered. 
We can express the sequence of trees formed as
\beq\label{eq:T_i}
T_i(t)=u_{v_i}\prod_{w\in T_i:\,w'=v_i}\int_{-\infty}^t T_w(\tau)\,d\tau,
\eeq
where in the product we consider the tree $T_i$ with the root $v_i$ having the successors $w$. 

Let us define the numbers, the divergence rates of the multipliers $u_{v_i}$,
\beqn
r_i\defas r_{v_i}.
\eeqn
Analogously to the numbers $s_v$ in \eqref{eq:n_v-def}, we set
\beqn
s_i\defas \sum_{w\in T_i:\,w'=v_i}\max{(0,n_w)}.
\eeqn
The numbers $n_w$ are the ones defined in $\eqref{eq:n_v-def}$ for the original tree, $T_0$. Notice that, although $s_0=s_{v_0}$, $s_i$ is not simply equal to $s_{v_i}$, but is the analogue in the tree $T_i$ of which $v_i$ is the \emph{root}. Similarly to \eqref{eq:s_v-bound}, $s_i$ bounds the divergence rate of the product in \eqref{eq:T_i}: if $\delta>0$,
\beq\label{eq:s_i-bound}
\biggl(\, \prod_{w\in T_i:\,w'=v_i}\int_{-\infty}^{t} T_w(\tau)\,d\tau\biggr) e^{-(s_i+\delta)\gamma t}\to 0\quad\text{as}\quad t\to\infty.
\eeq
The following, elementary, property is useful.
\begin{lemma}\label{lem:s+r}
For every $i$,
$
s_{i+1}+r_{i+1}\leq s_i.
$
\end{lemma}
\begin{proof}
Since $r_{i+1}\defas r_{v_{i+1}}$, $n_{v_{i+1}}\defas r_{v_{i+1}}+s_{v_{i+1}}$, and
$
\sum_{w\in T_{i}:w'=v_{i+1}}\max{(0,n_w)}=s_{v_{i+1}},
$
and
\beqn
s_{i+1}\defas\!\sum_{w\in T_{i+1} : w'=v_{i+1}}\max{(0,n_w)}=\!\!\sum_{w\in T_{i}:\,w\neq v_{i+1}, w'=v_{i}}\max{(0,n_w)}  +\!\!\sum_{w\in T_{i} : w'=v_{i+1}}\max{(0,n_w)},
\eeqn
we get
$
s_{i+1}+r_{i+1}=\sum_{w\in T_{i}:\,w\neq v_{i+1}, w'=v_{i}}\max{(0,n_w)}+n_{v_{i+1}}\leq s_i.
$
Here we used the fact that $v_{i+1}'=v_i$ in the tree $T_i$.
\end{proof}

Let $\tilde k_{0}\defas 0$ and $Q_0\defas 0$. Suppose that at the $i$th step we are considering the integral 
\beqn
\int_0^\infty e^{(iQ_i\cdot \omega+\tilde k_i\gamma-R)t}\,t^{p_i}u_{v_i}(ze^{\gamma t},q_{v_i})e^{iq_{v_i}\cdot \omega t}\,\prod_{w\in T_i:\,w'=v_i}\int_{-\infty}^{t} T_w(\tau;z,\tilde q_w)\,d\tau \,dt;
\eeqn
\textit{cf}.\@ \eqref{eq:integraali}. Then the integration-by-parts procedure described above takes us at the $(i+1)$st step to a similar integral with $i$ replaced by $i+1$, defining
\beq\label{eq:k-tilde-rec}
\tilde k_{i+1}\defas \tilde k_{i}+k_i\mathand Q_{i+1}\defas Q_i+q_{v_i},
\eeq
multiplied by the factor
\beq\label{eq:R-fraction}
\frac{z^{k_i}u_{v_i,k_i}(q_{v_i})}{{\bigl(R-\tilde k_{i+1}\gamma-i\omega\cdot Q_{i+1}\bigr)}^{1+p_i-p_{i+1}}}.
\eeq
The integer indices $p_{i+1}$ and $k_i$ can assume the values
\beq\label{eq:k-range}
0\leq p_{i+1}\leq p_i\leq 1\mathand k_i=-\tilde k_i-s_i,\dots,r_i.
\eeq
With the aid of Lemma~\ref{lem:s+r}, it is straightforward to see that 
\beq\label{eq:index-sum}
0\leq \tilde k_i+r_i+s_i\leq n_{v_0},
\eeq
where $n_{v_0}=s_{v_0}+r_{v_0}=s_0+r_0$ is the number describing the divergence rate of the original tree, $T_0$, in the sense of \eqref{eq:n_v-bound}. The number of possible values of $k_i$ above is thus at most $1+n_{v_0}$. This implies the two-sided (equivalent) bounds
\beqn
r_i-n_{v_0}\leq k_i\leq r_i \mathand 0\leq r_i-k_i\leq n_{v_0}.
\eeqn

We continue recursively until there are no nodes---alternatively, integrals that require regularization\linebreak[1]---left. The result is a sum with terms of two different species. We set
\beqn
R_i\defas R-\tilde k_{i}\gamma-i\omega\cdot Q_{i},
\eeqn
in order to make the presentation more compact, and also define (\textit{cf.} $E_v$ in \eqref{eq:int-by-parts})
\beqn
E_j(t)\defas E_j(t;z,\{\tilde q_w\,|\, w\in T_j, w'= v_j\})\defas  \!\!\prod_{w\in T_j:\,w'=v_j}\int_{-\infty}^{t} T_w(\tau;z,\tilde q_w)\,d\tau,
\eeqn
with the understanding that $E_{\abs{V_\text{int}}-1}\equiv 1$. The first class of terms is
\beq\label{eq:second-case}
\Biggl(\,\prod_{i=0}^{j-1}\frac{z^{k_i}u_{v_i,k_i}(q_{v_i})}{R_{i+1}^{1+p_i-p_{i+1}}}\Biggr) \int_0^\infty e^{(iq_{v_j}\cdot \omega-R_j) t} \,t^{p_j} \,u_{{v_j},<-\tilde k_j-s_j}(z e^{\gamma t},q_{v_j}) \,E_j(t) \,dt, 
\eeq
for $0<j< \abs{V_\text{int}}$. Notice that $s_{\abs{V_\text{int}}-1}=0$. Second, there are the terms ($0<j< \abs{V_\text{int}}$)
\beq\label{eq:third-case}
\Biggl(\,\prod_{i=0}^{j-1}\frac{z^{k_i}u_{v_i,k_i}(q_{v_i})}{R_{i+1}^{1+p_i-p_{i+1}}}\Biggr)\,\frac{z^{k_j}u_{v_j,k_j}(q_{v_j})}{R_{j+1}^{1+p_j}} \, E_j(0).
\eeq

\subsection{Estimates}\label{subsec:estimates}
By the definition of $s_j$, we can bound
\beqn
\abs{E_j(t)}\leq C_{E_j}e^{(s_j+\frac 14)\gamma t}e^{-\sigma\sum{\abs{\tilde q_w}}}\qquad(t\geq 0),
\eeqn  
for $\abs{z}=1$ with $\abs{\arg{z}}\leq \vartheta$. Deferring the proof to Appendix~\ref{app:computations}, we formulate
\begin{lemma}\label{lem:multiplier-bound}
The coefficients $u_{v_i,k_i}(q)$ satisfy
\beqn
\abs{u_{v_i,k_i}(q_{v_i})}\leq C_{v_i}\tau^{k_i-r_i} e^{-\sigma\abs{q_{v_i}}},
\eeqn 
where $\tau$ is as in Lemma~\ref{lem:r_v}. On $\{1\leq\abs{z}\leq \tau\inv\,\,\text{\rm with}\,\,\abs{\arg{z}}\leq\vartheta\}\bigcup \,\{\abs{z}\geq \tau\inv\}$, 
\beqn
\abs{u_{v_i,< -\tilde k_i-s_i}(z,q_{v_i})}\leq C_{v_i}(\tau\abs{z})^{-\tilde k_i-s_i-1} 2^{n_{v_0}} \tau^{-r_i} e^{-\sigma\abs{q_{v_i}}}.
\eeqn
When $\abs{z}=1, \abs{\arg{z}}\leq \vartheta$, and $\abs{\impart\theta}\leq\sigma$ it holds true, for suitable $\bar r_i\in\Z$, that
\beqn
\abs{u_{v_i}(ze^{\gamma t},\theta+\omega t)}\leq C_{v_i}
\begin{cases}
e^{\bar r_i \gamma t}, & t<0, \\
e^{r_i \gamma t}, & t\geq 0.
\end{cases}
\eeqn
\end{lemma}


Recalling \eqref{eq:index-sum}, the integral in \eqref{eq:second-case} is bounded by
$
(2\tau\inv)^{n_{v_0}+1}\frac{C_{v_j}C_{E_{j}}}{\gamma^{1+p_j}} e^{-\sigma\abs{q_{v_j}}-\sigma\sum\abs{\tilde q_w}},
$
when $\repart{R}\geq - \half\gamma$, $\abs{z}=1$, and $\abs{\arg{z}}\leq \vartheta$. Thus, the bounds on  \eqref{eq:second-case} and \eqref{eq:third-case} read
\beq \label{eq:second-case1}
\Biggl(\,\prod_{i=0}^{j-1}\frac{C_{v_i}}{\bigl|R_{i+1}^{1+p_i-p_{i+1}}\bigr|}\Biggr)\frac{C_{v_j}C_{E_{j}}}{\gamma^{1+p_j}}(2\tau\inv)^{n_{v_0}+1}\tau^{\sum_{i=0}^{j-1}(k_i-r_i)}e^{-\sigma\sum_{i=0}^j\abs{q_{v_i}}-\sigma\sum\abs{\tilde q_w}}
\eeq
and
\beq\label{eq:third-case1}
\Biggl(\,\prod_{i=0}^{j-1}\frac{C_{v_i}}{\bigl|R_{i+1}^{1+p_i-p_{i+1}}\bigr|}\Biggr)\frac{C_{v_j}C_{E_{j}}}{\bigl|R_{j+1}^{1+p_j}\bigr|}\tau^{\sum_{i=0}^{j}(k_i-r_i)}e^{-\sigma\sum_{i=0}^j\abs{q_{v_i}}-\sigma\sum\abs{\tilde q_w}} , 
\eeq
respectively. Here $e^{-\sigma\sum_{i=0}^j\abs{q_{v_i}}-\sigma\sum\abs{\tilde q_w}}\leq e^{-\sigma\abs{q}}$, because the tree is a Fourier transform with index $q$, and the $q_{v_i},\tilde q_w$ come from convolutions. The factors of $\tau$ are controlled by

\begin{lemma}
For all $j$,
$
\sum_{i=0}^j(k_i-r_i)\geq -n_{v_0}$. Moreover, $n_{v_0}\leq \operatorname{d}(T_{v_0})$
 (see \eqref{eq:ell-bound}).
\end{lemma}

\begin{proof}
From \eqref{eq:k-tilde-rec} and \eqref{eq:k-range}, we clearly have $\tilde k_{i+1}\geq -s_i$. Then,
\begin{align*}
\sum_{i=0}^j(k_i-r_i) & =\tilde k_{j+1}-\sum_{i=0}^j r_i\geq -s_j-r_j-\sum_{i=0}^{j-1} r_i 
& \geq -s_{j-1}-\sum_{i=0}^{j-1} r_i\geq\dots\geq -s_0-r_0=-n_{v_0},
\end{align*}
with the aid of Lemma~\ref{lem:s+r}. The estimate $n_{v_0}\leq \operatorname{d}(T_{v_0})$ follows easily from \eqref{eq:ell-bound} and \eqref{eq:xi-sum}.
\end{proof}

By the last bound of Lemma~\ref{lem:multiplier-bound} it is clear that $C_{E_j}=\bigl(\,\prod_{i=j+1}^{\abs{V_\text{int}}-1} C_{v_i} \bigr)\widetilde C_{E_j}$. In fact, each integral in $E_j$ produces a $\gamma\inv$ in the upper bound, and we get
$
\widetilde C_{E_j}\leq (C \gamma\inv )^{\abs{V_\text{int}}-j-1}.
$
Starting from the explicit expressions of the multipliers, we estimate
$
\prod_{i=0}^{\abs{V_\text{int}}-1} C_{v_i} \leq (C g)^{\abs{V}} (C \epsilon)^{\operatorname{d}(T_{v_0})}
$
with the aid of \eqref{eq:K-bounds} and \eqref{eq:a-prod}. Here $V$, defined above \eqref{eq:scalar-subtree}, is the set of all nodes including the black dot (\raisebox{3pt}{\TC*}) end nodes. Since $\abs{V_\text{int}}\leq \abs{V}\leq 2\operatorname{n}(T_{v_0})\leq 4\bigl(\operatorname{d}(T_{v_0})-1\bigr)$, 
\beqn
\Biggl(\,\prod_{i=0}^{j}C_{v_i}\Biggr) C_{E_j} \leq  g^{\abs{V}+1+j-\abs{V_\text{int}}} (C \epsilon)^{\operatorname{d}(T_{v_0})} 
 \qquad (0<j<\abs{V_\text{int}}).
\eeqn

Notice that the fraction $1/R_{i}$ is analytic on the domain where 
$
\abs{R}\neq \abs{\tilde k_{i}\gamma+i\omega\cdot Q_{i}}.
$
Thus, all the $1/R_i$ are analytic in the punctured neighbourhood of the origin
\beqn
0<\abs{R}\leq \rho\defas  \half \min \bigl\{\{\gamma \} \cup\{ \abs{\omega\cdot Q_i}\;|\;Q_i\neq 0\} \bigr\}.
\eeqn
On the circle $\abs{R}=\rho$,
$
\abs{R_{i}^{1+p}}\inv \leq  \rho^{-1-p},
$
provided $p\geq -1$.
According to \eqref{eq:k-range}, 
\beqn
\prod_{i=0}^{j-1}\biggl| \frac{1}{R_{i+1}^{1+p_i-p_{i+1}}} \biggr| \leq \rho^{-j-p_0+p_j} \qquad(0<j<\abs{V_\text{int}}).
\eeqn
Subsequently, summing over $j$ the bounds on \eqref{eq:second-case1} and \eqref{eq:third-case1} we get
\beqn
\sum_{j=1}^{\abs{V_\text{int}}-1} (C \epsilon)^{\operatorname{d}(T_{v_0})} g^{\abs{V}+1+j-\abs{V_\text{int}}}  \rho^{-j-1-p_0}  e^{-\sigma\abs{q}} 
\leq (C \epsilon)^{\operatorname{d}(T_{v_0})}   (g/\rho)^{\abs{V}+1}  g^{-1} e^{-\sigma\abs{q}} 
\defasr B(g)e^{-\sigma\abs{q}}.
\eeqn
At the $\ell$th order,  $\abs{Q_i}\leq \ell N$ for each $i$, and thus $\rho\geq \half \min(\gamma,  a (\ell N)^{-\nu})$.

From the above it is also clear that the order of the pole at $R=0$ in our integrals does not exceed $\abs{V_\text{int}}+1$. For the purposes of Lemma~\ref{lem:shift-of-contour}, we compute (using $g,\gamma\leq C$)
\beqn
\begin{split}
& B(g)\sum_{J=0}^{\abs{V_\text{int}}+1}\frac{1}{J!} \Bigl(\frac{\rho \vartheta}{g}\Bigr)^J \leq C B(g)  
\leq (C_N \epsilon)^{\operatorname{d}(T_{v_0})} g^{-1}  (\ell!)^{4\nu}.
\end{split}
\eeqn
There are at most 
$
(4\ell)!\leq 4^{4\ell}(\ell!)^4
$
orders in which we can exhaust all of the up to $4\ell$ lines of a tree contributing at order $\ell$. 

\subsection{Remaining integrals}
The integral over $\R_-$ in \eqref{eq:Fintegral} is simple, because the integrand satisfies the identity \eqref{eq:F-identity} and has the analyticity properties stated below that equation. In other words, we can separate the (possible) pole and the constant term from the rest: writing $F(X^u;0,z,\theta)\equiv F(z,\theta)$,
\beqn
F(X^u;t,z,\theta)=t^p z\inv e^{-\gamma t} F_{-1}(\theta+\omega t)+t^p F_{0}(\theta+\omega t)+t^p \delta_1 F(ze^{\gamma t},\theta+\omega t).
\eeqn 
Here $p=0,1$. Applying $\regint_{-\infty}^0$ on each term separately, we get that $\regint_{-\infty}^0  F(X^u;t,z,\theta)$ equals
\begin{align*}
(-1)^p\sum_{q'}\frac{z\inv e^{iq'\cdot\theta} \hat F_{-1}(q')}{(iq'\cdot\omega-\gamma)^{1+p}} 
 \,+\, (-1)^p\sum_{q'\neq 0}\frac{e^{iq'\cdot\theta} \hat F_{0}(q')}{(iq'\cdot\omega)^{1+p}} \,+\, \int_{-\infty}^0 t^p\delta_1F(ze^{\gamma t},\theta+\omega t)\,dt.
\end{align*}
These terms are small compared to to the large bounds obtained for the $\regint_0^\infty$ part above.

We also have to study the coefficients $c_{ij}^p$ appearing in \eqref{eq:as-expansion}. This is most conveniently done in terms of the representation \eqref{eq:c_ij}. Since $M^u=\order{\epsilon}$, only
\beqn
\regint_{-\infty}^{-t}\Bigl[f^u_{\psi\varphi}(B^uM^u)^l K_{ij}^p\Bigr](\tau)\,d\tau
\eeqn
with $l \leq \ell$ can contribute to $c_{ij}^p$ at order $\ell$. We only need to consider $t\geq 0$, as $\Upsilon(t)=\Upsilon(-t)$. 

The above integral consists in obvious shorthand notation, through \eqref{eq:Ksplit}, of $4^l$ terms like
{\small
\beqn
\regint_{-\infty}^{-t} (-t-\tau_l)^{p_{l+1}}(fK_l)(\tau_l)\regint_{-\infty}^{\tau_l}(\tau_l-\tau_{l-1})^{p_l} (\bar K_l M)(\tau_{l-1})\cdots K_1(\tau_1)\regint_{-\infty}^{\tau_1}(\tau_1-\tau_0)^{p_1}(\bar K_1 M)(\tau_{0})\tau_0^{p_0}K_0(\tau_0),
\eeqn
}%
where $p_{l+1}=0$. If $p_0=1$ (instead of 0), we use $\tau_0=(\tau_0-\tau_1)+\dots+(\tau_{l-1}-\tau_l)-(-t-\tau_l)-t$, getting $l+2$ terms of the original form except that $p_0=0$ and either there is a factor $t$ or precisely one change $p_i\mapsto p_i+1$ occurs. Due to $M=\order{\epsilon g^2}$, \eqref{eq:K-bounds} and \eqref{eq:a-prod}, each $K_i\bar K_i M$ produces a factor $CA\abs{\epsilon} g^{p_i+1}$ to the upper bound, whereas analyticity yields $e^{-\sigma\abs{q_i}}$ (which we prefer although at each order we reduce to trigonometric polynomials); 
$
\prod_{i=0}^l e^{-\sigma\abs{q_i}}\leq  e^{-\sigma\abs{q}}.
$

Last, the integral involves a total of  $\sum_{i=1}^{l+1}(1+p_i)\leq 2l+3$ denominators of the form $\tilde k_i \gamma + i\omega\cdot Q_i$, where $\tilde k_i\in\Z$ and $Q_i\defas \sum_{j=0}^i q_j$ with $Q_l=q\neq 0$. We bound these by
\beqn
\left( \frac{1}{\min\bigl\{\{\gamma\}\cup \{|\omega\cdot Q_i| \;|\; Q_i\neq 0\}\bigr\} }\right)^{\sum_{i=1}^{l+1}(1+p_i)} \leq \left(\max(\gamma\inv,a\inv (N\ell)^\nu) \right)^{\sum_{i=1}^{l+1}(1+p_i)}.
\eeqn
Altogether, the bounds above yield
$
C_N^\ell \abs{\epsilon}^l g^{-4} e^{-\sigma\abs{q}}(\ell!)^{2\nu}.
$

Skipping further details, this is the upper bound on the integral. It is smaller than what was derived for the integral $\regint_{-\infty}^\infty F$---and thus for $\E^p_{ij}$ in \eqref{eq:as-expansion}---above. In conclusion, among the contributions
$
\bigl(c_{ij}^p\bigr)^{\ell_1}\bigl(\E_{ij}^p\bigr)^{\ell_2}
$
($\ell_1+\ell_2=\ell-1$) to $\Upsilon^\ell$,  $\bigl(\E_{ij}^p\bigr)^{\ell-1}$ is the most dangerous one. 

\enlargethispage{5mm}
\begin{remark}
Above, the sums over $q_i$, with $\sum_i q_i=q$ and $\sum_i \abs{q_i}\leq \ell N$ were dealt with as follows. Since the analyticity domain with respect to $\theta$ is the compact $\{\abs{\impart{\theta}}\leq \sigma\}$, it can be substituted by some $\{\abs{\impart{\theta}}\leq \sigma'\}$ with $\sigma<\sigma'$. Then, for every $i$ we actually have the factor $e^{-\sigma'\abs{q_i}}$ in the estimates above. While $e^{-\sigma\sum_i\abs{q_i}}\leq e^{-\sigma\abs{q}}$, we get rid of the sums over $q_i$:
\beqn
\sum_{q_i}e^{-(\sigma'-\sigma)\abs{q_i}}\leq C_{\sigma'-\sigma}.
\eeqn
A similar remark applies to the Fourier indices in Subsection~\ref{subsec:estimates}. \qed
\end{remark}

\section{Discussion}

Each term in the asymptotic expansion \eqref{eq:as-expansion} of the splitting matrix $\Upsilon$ is proportional to an integral of the form $\regint_{-\infty}^\infty \de_\theta F(X^u;\tau,z,\theta)\,d\tau$. These we showed to be exponentially small in the limit $g\to 0$ at all orders $\ell$, by extending the integrands analytically into a wedge $\{\abs{\arg{z}}\leq \vartheta\}$ on the complex plane and then shifting the contour of integration. A key point is that we had to do this \emph{order by order}, because the series $\sum_{\ell=0}^\infty \epsilon^\ell X^{u,\ell}(z,\theta)$ is not expected to converge for large values of $\abs{z}$ if $\epsilon$ is fixed.

The large powers of the factorial $\ell!$, associated with the regularized integrals, are produced by accumulation of poles at the origin in the $R$ plane. To some extent the factorials are artifact, as is shown by the following simple example. In order to study, say, the integral
\beqn
\regint_0^{\infty}u(\theta+\omega t)\int_{-\infty}^t v(\theta+\omega \tau) ze^{\gamma\tau}\,d\tau\,dt,
\eeqn
we have to show that the sums
\beqn
\sum_p \hat u(p)\hat v(q-p)\int_0^{\infty}e^{(ip\cdot\omega-R) t}\int_{-\infty}^t e^{(i(q-p)\cdot \omega+\gamma) \tau} \,d\tau\,dt
\eeqn
extend analytically to a (punctured) neighbourhood of the origin, $R=0$. We integrate by parts, just as in \eqref{eq:int-by-parts} when extending analytically the tree integral \eqref{eq:reg-tree-int}, and get
\begin{align*}
\sum_p  \hat u(p)\hat v(q-p)\frac{1}{R-ip\cdot\omega}\Biggl\{\frac{1}{(i(q-p)\cdot \omega+\gamma)}+\frac{1}{R-(\gamma+iq\cdot\omega)}\Biggr\}.
\end{align*}
Here the pole at $ip\cdot \omega$ gets arbitrarily close to the origin, unless we restrict $p$ somehow---for instance, by considering trigonometric polynomials. Either by simplification, or by computing the same expression directly by starting from the inner integral, we obtain
\beqn
\sum_p\hat u(p) \hat v(q-p) \frac{1}{i(q-p)\cdot \omega+\gamma}\cdot \frac{1}{R-(\gamma+iq\cdot\omega)}.
\eeqn 
In the latter form there is no problem; the pole has cancelled. Of course, this is a naive example and in general it is hard to see whether a given pole popping out of the integration-by-parts procedure should really be there.

Also the coefficients $c_{ij}^p$ appearing in Proposition~\ref{prop:asymptotic} produced large powers of $\ell!$. Even though integration by parts was not exploited, the source of the factorials was again the accumulation of poles at the origin in the $R$ plane. In both this case and the previous, the type of ``divergence'' is very similar to what is encountered in KAM theory. There repeated resonances, or arbitrarily many occurrences of the operator $\D\inv$ in convolutions, ruin absolute convergence of the Fourier--Taylor expansion of a solution by producing high powers of the factorial $\ell!$. On the other hand, the state of affairs can be cured by well-known resummations, as in \cite{GallavottiTwistless}. Such resummations still escape us in the context of homoclinic splitting.

\appendix 
\section{Some computations and proofs}\label{app:computations}

\begin{proof}[Proof of Theorem~\ref{thm:splitmatrel}]
It is known that the whiskers are Lagrangian manifolds \cite{EliassonBiasymptotic,DelshamsPotential,LochakMemoirs}; they are graphs over the angles and, moreover, the actions are gradients of potentials. In other words, there exist generating funcions $S^{u,s}$ such that
\beqn
I^{u,s}\equiv \de_\phi S^{u,s}(\phi,\psi)\mathand A^{u,s}\equiv \de_\psi S^{u,s}(\phi,\psi).
\eeqn
We consider the latter row vectors. Differentiating both sides of 
\beqn
\Ham^{u,s}\defas\Ham(\phi,\psi,\de_\phi S^{u,s},\de_\psi S^{u,s})\equiv E
\eeqn
with respect to the $i$th component of $(\phi,\psi)\defasr\varphi=(\varphi_0,\varphi_1,\dots,\varphi_d)$ yields
\beq\label{eq:differentiated}
0=\Ham^{u,s}_i+\sum_{j=0}^d \Ham^{u,s}_{d+1+j}\,\de_{\varphi_i}\de_{\varphi_j}S^{u,s}=\Ham^{u,s}_i+\sum_{j=0}^d (\de_\varphi^2 S^{u,s})_{ij} \de_{J_j}\Ham^{u,s}.
\eeq
Here $J\defas(I,A)$ and $\de_{J_j}\Ham^{u,s}$ means $\Ham^{u,s}_{d+1+j}=\de_{J_j}\Ham(\phi,\psi,\de_\phi S^{u,s},\de_\psi S^{u,s})$.

From \eqref{eq:differentiated} it follows that, \emph{on the homoclinic trajectory},
\beq\label{eq:identity}
\de_J \Ham^{u,s}\,\de_\varphi^2(S^u-S^s)=0
\eeq
because one has $\de_\varphi S^u=\de_\varphi S^s$ and therefore also $\Ham^u_i=\Ham^s_i$ for $i=0,\dots,2d+1$. For our particular Hamiltonian, $\de_J \Ham^{u,s}=(I,A)^{u,s}$, where we now drop the superscripts $u,s$ and evaluate everything at the homoclinic point
\beq\label{eq:hp}
\varphi=(\phi,\psi)=(0,\omega t)+X^u(e^{\gamma t},\omega t) =(0,\omega t)+X^s(e^{\gamma t},\omega t)
\eeq 
below. The $\psi$ component of \eqref{eq:identity} reads
\beqn
I\,\de_\psi\de_\phi(S^u-S^s) + A\,\de^2_\psi(S^u-S^s) = 0,
\eeqn
or, recalling $I=I^0+\order{\epsilon}=2g/\cosh{gt}+\order{\epsilon}\neq 0$ (taking $\tilde \epsilon\defas g\inv e^{g\abs{t}}\epsilon$ small), 
\beq\label{eq:consequence}
\de_\psi\de_\phi(S^u-S^s)=-I\inv A \,\de^2_\psi(S^u-S^s).
\eeq
Here $A$ is the row vector $(A_1,\dots,A_d)$, such that both sides of the equality are row vectors. 

Next, let us perform the coordinate transformations 
$
(\phi,\psi)=F^{u,s}(z,\theta)\mapsto(z,\theta).
$
To this end, we observe that the splitting vector satisfies
\beqn
\Delta^T(\phi,\psi) =A^u-A^s = \de_\psi (S^u-S^s)(\phi,\psi).
\eeqn
Then, the $\theta$ derivative of the column vector $\Delta$ is the square matrix
\beqn
\de_\theta\Delta=\de_\psi\Delta\,\de_\theta\psi+\de_\phi\Delta\,\de_\theta\phi,
\eeqn
by the chain rule. In particular, by \eqref{eq:consequence},
\begin{align}\label{eq:derivative-relation}
\de_\phi\Delta &=  [\de_\psi\de_\phi(S^u-S^s)]^T = -\de^2_\psi(S^u-S^s)\,I\inv A^T  =-\de_\psi\Delta\,\,I\inv A^T
\end{align}
holds at a homoclinic point (see \eqref{eq:hp}), such that
\beqn
\de_\theta\Delta=\de_\psi\Delta\,\bigl(\de_\theta\psi - I\inv A^T\de_\theta\phi\bigr).
\eeqn
The matrix
$
M  \defas \de_\theta\psi - I\inv A^T\de_\theta\phi
$
has the asymptotic expression
\beqn
M =\bigl(\one+\order{\epsilon}\bigr)-\Bigl(\frac{2g}{\cosh{gt}}+\order\epsilon\Bigr)\inv \cdot \bigl(\omega+\order{\epsilon}\bigr)^T \cdot \order{\epsilon}
\eeqn
with $\epsilon= ge^{-g\abs{t}}\tilde\epsilon$, as before, and $\tilde\epsilon$ small.
\end{proof}

\begin{proof}[Proof of Lemma~\ref{lem:Lazutkin}. (Adapted from \cite{Sauzin}).]
The Fourier transform of \eqref{eq:Lazutkin-id} yields
\beq\label{eq:Lazutkin-Fourier}
\hat F(s,q)=\hat F(0,q)\, e^{-i(g\inv \omega\cdot q) s},
\eeq
which is entire in $s$. But
$
\abs{\hat F(0,q)}\,e^{(g\inv \omega\cdot q)\,\impart s}=\abs{\hat F(s,q)}\leq B(g) e^{-\eta\abs{q}}
$
for $s\in [-i\vartheta,i\vartheta]$, and
\beqn
\abs{\hat F(0,q)}\leq B(g) e^{-\vartheta g\inv \abs{\omega\cdot q}-\eta\abs{q}}.
\eeqn
Finally, plugging this into \eqref{eq:Lazutkin-Fourier}, we get
\beqn
\abs{\hat F(s,q)}\leq B(g) e^{-(\vartheta-\abs{\impart{s}})g\inv \abs{\omega\cdot q}-\eta\abs{q}}\qquad(s\in\C).
\eeqn
For $\abs{\impart{s}}\leq \vartheta$ and $\abs{\impart{\theta}}\leq\eta'<\eta$, the series $\sum_{q\in\Z^d}\hat F(s,q)e^{iq\cdot \theta}$ is uniformly convergent and, as such, provides the analytic extension.

Since $\alpha x^{-\nu}+\beta x\geq \alpha(\nu+1)\bigl(\frac{\alpha\nu}{\beta}\bigr)^{-\nu/(\nu+1)}$ for positive $\alpha$, $\beta$, $\nu$, and $x$, we get from the Diophantine condition \eqref{Dio} that
\beqn
\abs{\hat F(s,q)}\leq B(g) e^{-\vartheta g\inv a\abs{q}^{-\nu}-\eta\abs{q}}\leq B(g) e^{-\delta\abs{q}}e^{-w(\vartheta,\eta-\delta)g^{-1/(\nu+1)}}
\eeqn
holds if $s\in\R,\,q\in\Z^d\nonzero$, $0<\delta<\eta$, and $w(\vartheta,\eta-\delta)\defas (\vartheta a)^{1/(\nu+1)}(\eta-\delta)^{\nu/(\nu+1)}(\nu+1)\nu^{-\nu/(\nu+1)}$. Moreover,
$
\sum_{q\in\Z^d\nonzero} e^{-\delta\abs{q}}\leq C\delta^{-d},
$
where $C$ only depends on the dimension $d$. 

By \eqref{eq:Lazutkin-Fourier}, $\widetilde F\defas\average{F(s,\piste)}=\hat F(s,0)=\hat F(0,0)$, such that $\widetilde F$ does not depend on $(s,\theta)$.
\end{proof}

\begin{proof}[Proof of Lemma~\ref{lem:shift-of-contour}]
By shifting the contour of integration from $\R$ to the complex plane by $i t_q\defas i\sgn(\omega\cdot q)\vartheta g\inv$ units, we compute
\begin{align*}
\regint_{-\infty}^\infty  t^p \hat h(e^{gt},q) e^{i q\cdot \omega t}\,dt &=\res_{R=0}\frac{1}{R}\int_{-\infty}^\infty e^{-R\abs{t}} t^p \hat h(e^{gt},q) e^{i q\cdot \omega t}\,dt \\
&=\res_{R=0}\frac{1}{R}\Bigl\{e^{-iR t_q}H_q(R)+e^{iR t_q}I_q(R)\Bigr\} e^{-\vartheta g\inv\abs{\omega\cdot q}},
\end{align*}
where $H_q$ is defined in \eqref{eq:H_q} and 
$
I_q(R)\defas\int_{-\infty}^0 e^{(iq\cdot\omega+R)t}(t+it_q)^p\hat h(e^{g(t+it_q)},q)\,dt.
$
There are, \textit{a priori}, two additional line integrals $\int_0^{it_q}$, but they cancel due to the residue at $R=0$, as is easily checked. Because $\hat h(\piste,0)=0$, $I_q(R)$ does not have a pole at $R=0$. Hence,
\beqn
\res_{R=0}\frac{e^{iR t_q}I_q(R)}{R}=\regint_{-\infty}^0 e^{iq\cdot\omega t}(t+it_q)^p\hat h(e^{g(t+it_q)},q)\,dt.
\eeqn
If $H_q(R)$ has a pole of order $k$ at $R=0$, then
\beqn
\biggl|\res_{R=0}\frac{e^{-iR t_q}H_q(R)}{R}\biggr|=\biggl|\sum_{j=0}^k \frac{(-it_q)^j H_{q,-j}}{j!}\biggr|\leq \sum_{j=0}^k \frac{1}{j!}\Bigl(\frac{\rho\vartheta}{g}\Bigr)^j\sup_{\abs{R}=\rho}\abs{H_q(R)},
\eeqn
because the Laurent coefficients
$
H_{q,-j}\defas \frac{1}{2\pi}\oint_{\abs{R}=\rho}\frac{H_q(R)}{R^{-j+1}}\,dR
$
can be bounded from above by
$
\abs{H_{q,-j}}\leq \rho^j\sup_{\abs{R}=\rho}\abs{H_q(R)},
$
whenever the circle $\abs{R}=\rho$ is inside the domain of $H_q$. 

Under the assumptions of the lemma, for any $0<\delta<\sigma$ and $q\in\Z^d\nonzero$,
\beqn
\biggl|\,\regint_{-\infty}^\infty  t^p \hat h(e^{gt},q) e^{i q\cdot \omega t}\,dt \biggr|\leq \biggl[A(g)+B(g)\sum_{j=0}^k \frac{1}{j!}\Bigl(\frac{\rho\vartheta}{g}\Bigr)^j\biggr]e^{-\delta\abs{q}}e^{-w(\vartheta,\sigma-\delta)g^{-1/(\nu+1)}},
\eeqn
by mimicking the proof of Lemma~\ref{lem:Lazutkin}. Here $w(\vartheta,\sigma-\delta)\defas (\vartheta a)^{1/(\nu+1)}\bigl(\frac{\sigma-\delta}{\nu}\bigr)^{\nu/(\nu+1)}(\nu+1)$. Summation over $q$ produces a factor $C\delta^{-d}$.
\end{proof}

\begin{proof}[Proof of Lemma~\ref{lem:multiplier-bound}]
Here we usually omit the subindices $i$, $v$, and $v_0$. According to Lemma~\ref{lem:r_v}, $u_v$ is analytic on $\{ \abs{z} \geq \tau\inv \} \times \{ \abs{\impart{\theta}}\leq \sigma \}$ with a pole of order $r$ at $z=\infty$. A Cauchy estimate then reads
$
\abs{u_{v,k}(\theta)}\leq \tau^k \sup_{\abs{z}=\tau\inv}\abs{u_v(z,\theta)}\leq C \tau^{k-r}.
$
Now the first bound follows from $\abs{u_{v,k}(q)}\leq e^{-\sigma\abs{q}}\sup_{\abs{\impart{\theta}}\leq\sigma}\abs{u_{v,k}(\theta)}$.

The second inequality is a trivial consequence of the first one for $\abs{z}>2\tau\inv$. For the other values of $z$, one uses \eqref{eq:multiplier-split} and \eqref{eq:index-sum} as well as $\abs{z}^{-r}\abs{u_v(z,q)}\leq C e^{-\sigma\abs{q}}$ to bound 
\begin{align*}
\abs{u_{v,< -\tilde k-s}(z,q)} & \leq \abs{z}^{-\tilde k-s-1}\Bigl\{\abs{z}^{\tilde k+s+1} \abs{u_v(z,q)}+ C e^{-\sigma\abs{q}} \tau^{-\tilde k-s-1} \sum_{l=0}^{r+\tilde k+s} \abs{z\tau}^{l+1} \Bigr\} \\
& \leq C e^{-\sigma\abs{q}} \abs{z}^{-\tilde k-s-1} \Bigl\{(2\tau\inv)^{\tilde k+s+r+1} + \tau^{-\tilde k-s-r-1} 2^{r+\tilde k+s} \Bigr\}. 
\end{align*}
The last inequality in the lemma is obvious and is stated for completeness. 
\end{proof}%


\bibliographystyle{amsalpha}
\bibliography{References}

\end{document}